\documentclass[twocolumn,aip,jap,showpacs,10pt,superscriptaddress]{revtex4-1}

\usepackage{bm}
\usepackage{amssymb}
\usepackage{amsmath}
\usepackage{bm}
\usepackage{multirow}
\usepackage{graphicx}
\usepackage[usenames,dvipsnames,svgnames]{xcolor}
\usepackage{hyperref}
\usepackage[normalem]{ulem}
\hypersetup{
pdfnewwindow=true,      
colorlinks=true,        
linkcolor=Blue,         
citecolor=Blue,         
filecolor=Blue,         
urlcolor=Blue           
}

\begin{document}

\title{Heat transport across graphene/hexagonal-BN tilted grain boundaries from phase-field crystal model and molecular dynamics simulations}
\author{Haikuan Dong}
\affiliation{Beijing Advanced Innovation Center for Materials Genome Engineering, Corrosion and Protection Center, University of Science and Technology Beijing, Beijing, 100083, China}
\affiliation{MSP group, QTF Centre of Excellence, Department of Applied Physics, Aalto University, FI-00076 Aalto, Finland}
\affiliation{College of Physical Science and Technology, Bohai University, Jinzhou, 121013, China}
\author{Petri Hirvonen}
\affiliation{MSP group, QTF Centre of Excellence, Department of Applied Physics, Aalto University, FI-00076 Aalto, Finland}
\author{Zheyong Fan}
\email{brucenju@gmail.com}
\affiliation{MSP group, QTF Centre of Excellence, Department of Applied Physics, Aalto University, FI-00076 Aalto, Finland}
\affiliation{College of Physical Science and Technology, Bohai University, Jinzhou, 121013, China}
\author{Ping Qian}
\email{qianping@ustb.edu.cn}
\affiliation{Beijing Advanced Innovation Center for Materials Genome Engineering, Department of Physics, University of Science and Technology Beijing, Beijing, 100083, China}
\author{Yanjing Su}
\email{yjsu@ustb.edu.cn}
\affiliation{Beijing Advanced Innovation Center for Materials Genome Engineering, Corrosion and Protection Center, University of Science and Technology Beijing, Beijing, 100083, China}
\author{Tapio Ala-Nissila}
\affiliation{MSP group, QTF Centre of Excellence, Department of Applied Physics, Aalto University, FI-00076 Aalto, Finland}
\affiliation{Interdisciplinary Centre for Mathematical Modelling, Department of Mathematical Sciences, Loughborough University, Loughborough, Leicestershire LE11 3TU, UK}

\date{\today}

\begin{abstract}
We study the interfacial thermal conductance of grain boundaries (GBs) between monolayer graphene and hexagonal boron nitride (h-BN) sheets using a combined atomistic approach. First, realistic samples containing graphene/h-BN GBs with different tilt angles are generated using the phase-field crystal (PFC) model developed recently [P. Hirvonen \textit{et al.}, Phys. Rev. B \textbf{100}, 165412 (2019)] that captures slow diffusive relaxation inaccessible to molecular dynamics (MD) simulations. Then, large-scale MD simulations using the efficient GPUMD package are performed to assess heat transport and rectification properties across the GBs. We find that lattice mismatch between the graphene and h-BN sheets plays a less important role in determining the interfacial thermal conductance as compared to the tilt angle. In addition, we find no significant thermal rectification effects for these GBs.
\end{abstract}

\maketitle

\section{Introduction}

After the discovery of atomically thin two-dimensional (2D) graphene and other low-dimensional materials, there is a major effort in  constructing heterostructures made of them, either by stacking 2D materials vertically to form multi-layer heterostructures \cite{Novoselov2016science} or by stitching them laterally to form a 2D sheet with in-plane junctions. Graphene and hexagonal boron nitride (h-BN) are promising candidates for stable heterostructures as they have similar crystal structures with a lattice constant difference of $2\%$ only. To this end, graphene/h-BN in-plane heterostructures have already been fabricated and shown to be promising materials for the next-generation nanodevices \cite{ci2010atomic,Levendorf2012nature,liu2013plane,Han2013acsnano,gao2013nl,liu2014nl,Ling2016am}.

One of the most important properties in nanodevices is efficient thermal management in the context of nanophononics \cite{volz2016epjp}, which requires detailed microscopic understanding on the heat transport properties. To our best knowledge, thermal transport properties of graphene/h-BN grain boundaries (GBs) have not been studied experimentally but only theoretically using the atomistic Green's function \cite{ong2016prb} and molecular dynamics (MD) simulations \cite{liu2016nl}. Other works have also considered the case of multilayer graphene/h-BN GBs \cite{liang2020ijhmt} and the coexistence of other defects \cite{li2018jpcc,song2020pccp,wu2021jpcc} or disorder \cite{ni2019ijhmt} in addition to those in the GB.  However, the previous theoretical studies have considered a limited set of simple graphene/h-BN interfaces which do not fully represent all the possible structures of interest.

To gain a thorough microscopic understanding of the role of phonons in heat transport in lateral graphene/h-BN heterostructures MD simulations are the tool of choice. However, an efficient model construction method for the GBs has to be developed because the GBs develop and relax at diffusive time scales that may be well beyond what can be achieved with MD. In a previous work \cite{hirvonen2019prb} some of the present authors have developed an efficient and flexible phase-field crystal (PFC) model \cite{elder2002prl,elder2004pre} to describe the relaxed atomic configurations of multiple atomic species and phases coexisting in the same physical domain. Specifically, a PFC model for lateral graphene/h-BN heterostructures was constructed. \cite{hirvonen2019prb} The PFC models are a family of continuum methods which can model the atomistic structures and energetics of crystals, and their evolution at diffusive time scales that are usually inaccessible to atomistic MD simulations. PFC models for single-component systems such as graphene \cite{hirvonen2016prb} and binary systems such as h-BN \cite{taha2017prl} have been successfully applied to construct large and realistic bicrystalline and polycrystalline systems for further atomistic calculations such as quantum-mechanical density functional ones or classical MD simulations  \cite{fan2017nl,azizi2017cb,dong2018pccp}.

In this work, we use the PFC model \cite{hirvonen2019prb} to construct realistic bicrystalline systems consisting of graphene/h-BN GBs. We consider both lattice-matched and lattice-mismatched systems, both with a series of tilt angles between the graphene and h-BN sheets. After obtaining relaxed atomistic configurations for such systems, we use them as input to MD simulations and systematically study the heat transport properties across these graphene/h-BN GBs. We find that lattice mismatch between the graphene and h-BN sheets does not play a crucial role in determining the interfacial thermal conductance and no significant thermal rectification can be found. These findings suggest that heat transport properties in lateral graphene/h-BN heterostructures are not very sensitive to the actual microstructures of the GBs.

\section{Models and Methods}

\subsection{Realistic heterogeneous bicrystalline samples from a PFC model}

\begin{figure}[htb]
\begin{center}
\includegraphics[width=\columnwidth]{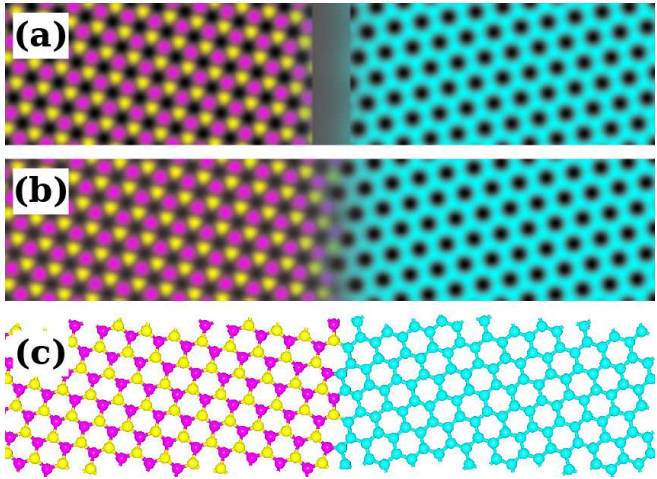}
\caption{Sample preparation using PFC. (a) A blow-up of the initial state with symmetrically tilted crystals of h-BN (magenta and yellow) and graphene (cyan) and with a narrow band of undercooled liquid between them. (b) The same region after PFC relaxation. (c) The atomistic configuration extracted from the relaxed PFC system.}
\label{fig-sample}
\end{center}
\end{figure}

PFC models are a family of continuum methods for studying polycrystalline systems across atomic-to-mesoscopic length scales. Conventional PFC models omit the fast phononic vibrations in favor of slow, diffusive dynamics \cite{elder2002prl}. This allows efficient relaxation of defected structures. We use here a PFC model which allows controlled phase separation and enables modeling heterostructures readily with different elemental compositions, lattice structures and elastic properties between the different phases. Full details and parameters of the model are given in Ref. \onlinecite{hirvonen2019prb}, but we outline the essentials here. We model graphene/h-BN using $N = 3$ periodic, smooth density fields $n_1$, $n_2$ and $n_3$ for carbon, boron and nitrogen, respectively. We initialize $n_i$ in a two-dimensional, periodic computational unit cell with two symmetrically tilted crystals and narrow bands of undercooled liquid between them (cf. Fig. \ref{fig-sample} (a)). We then assume the following dynamics and equilibrate $n_i$ numerically using a semi-implicit spectral method \cite{ref-pf-book}:

\begin{equation}
    \label{eq-pfc-dynamics}
    \begin{split}
        \frac{\partial n_i}{\partial t}
        \\
        = \nabla^2 \left( \vphantom{\sum_j^N} \alpha_{i} n_i + \beta_i \left( \nu_i^2 + \nabla^2 \right)^2 n_i + \gamma_i n_i^2 + \delta_i n_i^3 \right.
        \\
        \left. + \sum_{\substack{j = 1 \\ j \neq i}}^N \left( \vphantom{\frac{\gamma_j}{2}} \alpha_{ij} n_j + \beta_{ij} \left( \nu_{ij}^2 + \nabla^2 \right)^2 n_j \right. \right.
        \\
    \left. \left. + \frac{\gamma_{ij}}{2} \left( 2 n_i n_j + n_j^2 \right) + \epsilon_{ij} G \ast \left( G \ast n_j \right) \right) \vphantom{\sum_j^N} \right).
    \end{split}
\end{equation}

In the expression above, the first four terms are typical to single-component PFC models whereas the next four couple $n_i$ together. The last term is responsible for the phase separation and involves a Gaussian convolution (denoted by an asterisk operation) of a smoothed density $\eta_j$ where the atomic scale structures have been filtered out. The relaxed density fields [Fig. \ref{fig-sample} (b)] are converted into atomic coordinates [Fig. \ref{fig-sample} (c)] using a simple extension of the method in Refs. \onlinecite{hirvonen2016prb, ref-hirvonen-thesis}.

\begin{figure*}[htb]
\begin{center}
\includegraphics[width=2\columnwidth]{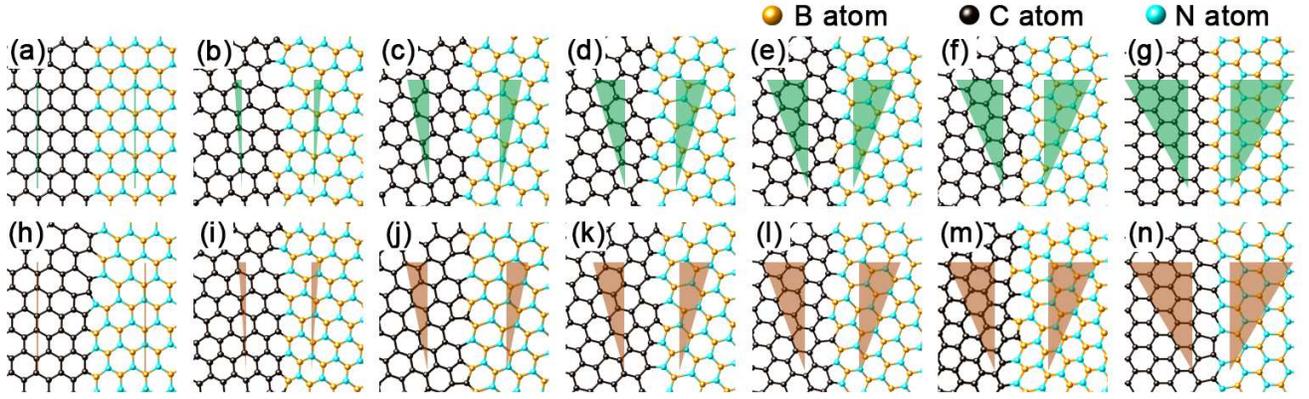}
\caption{The symmetrically tilted graphene/h-BN GBs. (a)-(g): The series of lattice-matched-GBs with increasing tilt angles; (h)-(n): The series of lattice-mismatched-GBs with increasing tilt angles. For both the upper and lower panels, the tilt angles $2\theta$ (indicated by the two wedges in each panel) from left to right are respectively $0 ^\circ$, $9.43 ^\circ$, $21.79 ^\circ$, $32.20 ^\circ$, $42.10 ^\circ$, $46.83 ^\circ$, and $60 ^\circ$.}
\label{figure:grain_boundaries}
\end{center}
\end{figure*}

Figure \ref{figure:grain_boundaries} illustrates the local structures of the graphene/h-BN GBs from the PFC model. The structures in the upper panels are obtained by assuming a common lattice constant of $2.46$ \AA~ for graphene and h-BN in the PFC model, and are called lattice-matched GBs. The structures in the lower panels are obtained by assuming a lattice constant of $2.46$ \AA~ for graphene and a lattice constant of $2.51$ \AA~ for h-BN in the PFC model, and are called lattice-mismatched GBs. Previous works \cite{ong2016prb,liu2016nl,liang2020ijhmt, wu2021jpcc} have only considered graphene/h-BN GBs with a tilt angle of $0^\circ$ (armchair-oriented GB) or $60 ^\circ$ (zigzag-oriented GB). For these two special tilt angles, the lattice-matched GBs do not contain any topological defects, see Figs. \ref{figure:grain_boundaries}(a) and \ref{figure:grain_boundaries}(g), while the lattice-mismatched GBs contain sparsely distributed $5-7$ defects, see Figs. \ref{figure:grain_boundaries}(h) and \ref{figure:grain_boundaries}(n).

All samples from our PFC model consist of a graphene sheet and a h-BN sheet connected by two GBs and are periodic in the planar $x$ and $y$ directions. The GBs align along the $y$ direction and we are interested in heat transport in the $x$ direction, perpendicular to the GBs. Both the graphene and the h-BN sheets have a length of about $100$ nm in the $x$ direction. For the lattice-matched-GBs, the widths in the $y$ direction are about $10$ nm. For the lattice-mismatched-GBs, it is harder to find a periodic cell with small width, and the widths in the $y$ direction vary from about $14$ nm to about $170$ nm depending on the tilt angle.

\begin{figure}[htb]
\begin{center}
\includegraphics[width=\columnwidth]{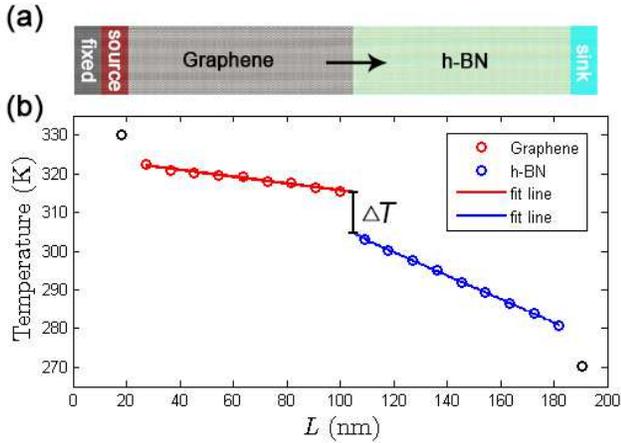}
\caption{(a) A schematic illustration of the NEMD simulation setup, where heat flows from the graphene sheet to the h-BN sheet. (b) The steady-state temperature profile. The temperature jump $\Delta T$ can be obtained by linear fits to the temperature profiles on both sides of the GB.}
\label{figure:nemd_example}
\end{center}
\end{figure}

\subsection{Heat transport properties form NEMD simulations}

With the bicrystalline samples available, we proceed to study the heat transport properties. To this end, we use the open-source \textsc{gpumd} package \cite{fan2013cpc,fan2017cpc} to perform NEMD simulations. The Tersoff many-body potential \cite{Tersoff1989prb} constructed by Kinaci \textit{et al.} \cite{kinaci2012prb} is employed to describe the interatomic interactions. The \textsc{gpumd} package is fully implemented in graphics processing units (GPU) and the computational speed with one Nvidia Tesla V100 for the MD simulations with the Tersoff potential is about $10^8$ atom-step per second. This high computational efficiency allows us to perform extensive MD simulations to characterize the heat transport properties of the graphene/h-BN heterostructures.

\subsubsection{The NEMD simulation protocol}

Referring to Fig. \ref{figure:nemd_example}(a), we first consider heat transport from the graphene side to the h-BN side. Later, we will also consider the opposite direction in the context of thermal rectification. Due to the use periodic boundary conditions, we only need to freeze a single block of atoms (marked as ``fixed'') between the source and sink regions to realize an insulating wall between them. Then heat will only flow along the direction as indicated by the arrow.

In the NEMD simulation of a sample, we first equilibrate the system at $10$ K for $25$ ps, then linearly heat up the system from $10$ K to $300$ K for $25$ ps, and then equilibrate the system at $300$ K for $25$ ps. In the stages above, the NPT ensemble with a Berendsen thermostat and barostat \cite{Berendsen1984jcp} is used. After equilibraration, we remove the global thermostat and barostat and apply local thermostats to the heat source and sink regions. In this regard, we follow the established practice \cite{li2019jcp,hu2020prb} by using the Langevin thermostat \cite{Bussi2007pre}. After achieving a steady state, the heat source and sink regions will have higher and lower temperatures which are chosen as $330$ K and $270$ K, respectively. All the systems reach a steady state within $500$ ps. After this, we use another $500$ ps to sample the temperature profile. In all the MD simulations a time step of $0.25$ fs is used which is small enough. To ensure high statistical accuracy, we perform three independent runs for each system.

A typical temperature profile is shown in Fig. \ref{figure:nemd_example}(b). We see that there are temperature jumps at three places: between the heat source region and the graphene sheet, around the GB, and between the heat sink region and the h-BN sheet. The temperature jumps around the heat source and sink are related to ballistic contact resistance, which exist even in short crystalline systems \cite{li2019jcp,hu2020prb}. The temperature jump $\Delta T$ around the GB, on the other hand, is the hallmark of the interfacial Kapitza thermal resistance $R$, or the interfacial Kapitza thermal conductance $G$ defined as (the superscript ``c'' means ``classical'', as will be further discussed below)
\begin{equation}
\label{equation:resistance}
G^{\rm c} = \frac{1}{R} = \frac{Q}{\Delta T}.
\end{equation}
Here,
\begin{equation}
\label{equation:q}
Q = \frac{1}{V}\sum_{i\in V}
\left\langle
\bm{W}_i \cdot \bm{v}_i
\right\rangle_{\rm ne}
\end{equation}
is the nonequilibrium ensemble average (hence the subscript ``ne'') of the heat flux \cite{fan2015prb,fan2019prb,Gabourie2021prb} in the transport direction ($x$ direction) determined in the steady state for a part of the system containing the GB, where $V$ is the control volume of this part. To calculate $V$, a thickness of $0.335$ nm is assumed for the single layer. In Eq. (\ref{equation:q}), $\bm{v}_i$ is the velocity of atom $i$ and
\begin{equation}
\label{equation:virial}
\bm{W}_i = \sum_{j\neq i} x_{ij}
\frac{\partial U_j}{\partial \bm{r}_{ji}}
\end{equation}
is a ``vector'' formed by three components of the virial tensor of atom $i$, where $U_j$ is the potential energy of atom $j$ and $\bm{r}_{ij}=\bm{r}_{j}-\bm{r}_{i}$, $\bm{r}_i$ being the position of atom $i$. The summation index $i$ in Eq. (\ref{equation:q}) runs over the atoms in the volume $V$ and the summation index $j$ in Eq. (\ref{equation:virial}) runs over the neighbors of atom $i$. The total heat flux calculated using Eq. (\ref{equation:q}) is equal to that calculated based on the energy exchange rate within the heat source and sink regions under the action of the local thermostats, as has been demonstrated in previous works. \cite{fan2017prb,xu2018msmse}

\subsubsection{Spectral decomposition and quantum correction}

In the context of NEMD simulations, a spectral decomposition method \cite{Chalopin2013apl,saaskilahti2014prb,saaskilahti2015prb,zhou2015prb,fan2017prb} has been developed based on the force-velocity time-correlation function. This method has been later reformulated \cite{fan2019prb,Gabourie2021prb} into a more convenient form based on the virial-velocity time-correlation function. In this method, the thermal conductance can be integrated with respect to the phonon frequency $\omega$,
\begin{equation}
\label{equation:g}
G^{\rm c} =
\int_{0}^{\infty} \frac{d\omega}{2\pi}G^{\rm c}(\omega),
\end{equation}
where $G^{\rm c}(\omega)$ is called the spectral thermal conductance that can be calculated \cite{fan2019prb,Gabourie2021prb} from the following Fourier transform:
\begin{equation}
\label{equation:g_omega}
G^{\rm c}(\omega) =
\frac{2}{\Delta T}\int_{-\infty}^{+\infty} dt e^{i\omega t}  K(t).
\end{equation}
Here $K(t)$ is the virial-velocity time-correlation function defined as
\begin{equation}
K(t) = \frac{1}{V}\sum_{i\in V}
\left\langle
\bm{W}_{i}(0) \cdot \bm{v}_i(t)
\right\rangle_{\rm ne}.
\end{equation}
This formalism applies to general many-body interatomic potentials, including machine learning ones. \cite{fan2021arxiv}

The spectral thermal conductance $G^{\rm c}(\omega)$ defined in Eq. (\ref{equation:g_omega}) should be understood as the classical one. Quantum correction to $G^{\rm c}(\omega)$ has been shown to be feasible \cite{lv2016njp,saaskilahti2016aipa,azizi2017cb} by using the quantum modal heat capacity, which amounts to multiplying $G^{\rm c}(\omega)$ by a factor (the superscript ``q'' means ``quantum''),
\begin{equation}
G^{\rm q}(\omega) = G^{\rm c}(\omega) \frac{x^2e^x}{(e^x-1)^2},
\end{equation}
where $x = \hbar\omega/k_{\rm B}T$, and $\hbar$, $k_{\rm B}$, and $T$ are the reduced Planck constant, the Boltzmann constant, and the temperature, respectively. The total quantum-corrected interfacial thermal conductance is obtained by an integral with respect to the frequency:
\begin{equation}
\label{equation:g_quantum}
G^{\rm q} =
\int_{0}^{\infty} \frac{d\omega}{2\pi}G^{\rm q}(\omega).
\end{equation}

\section{Results and Discussion}

\subsection{Kapitza thermal conductance}

\begin{figure}[htb]
\begin{center}
\includegraphics[width=\columnwidth]{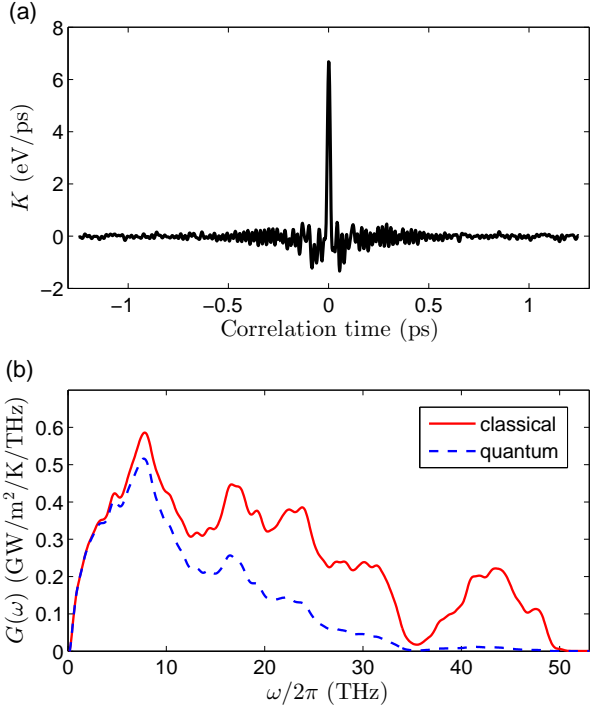}
\caption{(a) The virial-velocity time-correlation function. (b) The classical (solid line) and quantum (dashed line) spectral interfacial thermal conductance. The system considered here corresponds to the lattice-mismatched GB with $2\theta = 0^{\circ}$.}
\label{figure:spectral}
\end{center}
\end{figure}

Figure \ref{figure:spectral}(a) shows the virial-velocity time-correlation function $K(t)$ for the lattice-mismatched GB with zero tilt angle. The corresponding classical and quantum spectral interfacial thermal conductances $G^{\rm c}(\omega)$ and $G^{\rm q}(\omega)$ are respectively shown as the solid and dashed lines in Fig. \ref{figure:spectral}(b). It is important to consider both the positive and negative correlation times as shown in Fig. \ref{figure:spectral}(a) to calculate the spectral thermal conductance via the Fourier transform; using either the positive or the negative part only, i.e., assuming $K(-t)=K(t)$, could result in noticeable negative values in the spectral thermal conductance at particular frequencies. An important observation here is that the quantum correction strongly suppresses the contributions from the high-frequency modes.

\begin{figure}[htb]
\begin{center}
\includegraphics[width=\columnwidth]{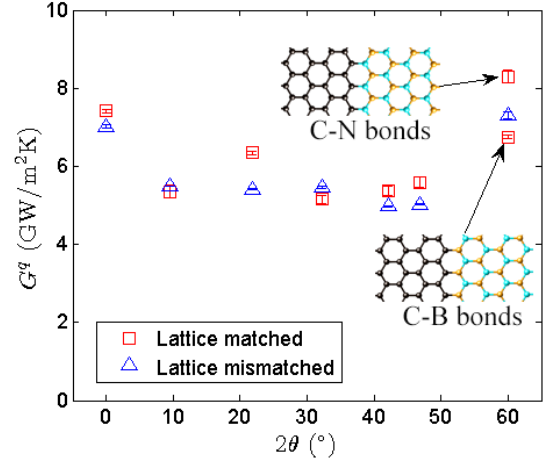}
\caption{Quantum-corrected interfacial thermal conductance $G^{\rm q}$ of lattice-matched (squares) and lattice-mismatched (triangles) GBs. In the case of matching lattice and $2\theta=60^{\circ}$, we consider GBs stitched with either C-N or C-B bonds.}
\label{figure:quantum_conductance}
\end{center}
\end{figure}

By integrating $G^{\rm q}(\omega)$ with respect to the frequency, we obtain the total quantum interfacial thermal conductance $G^{\rm q}$ for all the systems considered in this work, which are shown in Fig. \ref{figure:quantum_conductance}. There are a few important observations from Fig. \ref{figure:quantum_conductance}, as detailed below.

First, there is a strong tilt-angle dependence of the interfacial thermal conductance for both the lattice-matched and lattice-mismatched GBs. In both series, the largest conductance occurs at $2\theta=0^{\circ}$ or $60^{\circ}$, which is about $50\%$ larger than the smallest conductance occurring at $2\theta=32.20^{\circ}$ or $2\theta=42.10^{\circ}$. This general trend can be understood as there are more defects (hence higher GB energy) in the GBs with intermediate tilt angles. Similar trend exists in other systems such as twist grain boundaries in silicon. \cite{Schelling2004jap}

Second, the lattice-matched GB with $2\theta=21.79^{\circ}$ has an abnormally large thermal conductance compared to the lattice-matched GBs with nearby tilt angles. This abnormality also exists in pure graphene GB \cite{azizi2017cb} and pure h-BN GB \cite{dong2018pccp} with the same tilt angle. At this particular tilt angle, the lattice-matched GB consists of regularly arranged 5-7 rings and the resulting monolayer around the GB is quite flat. \cite{azizi2017cb} The flatness helps to enhance the phonon transmission across the GB.

\begin{figure*}[htb]
\begin{center}
\includegraphics[width=2\columnwidth]{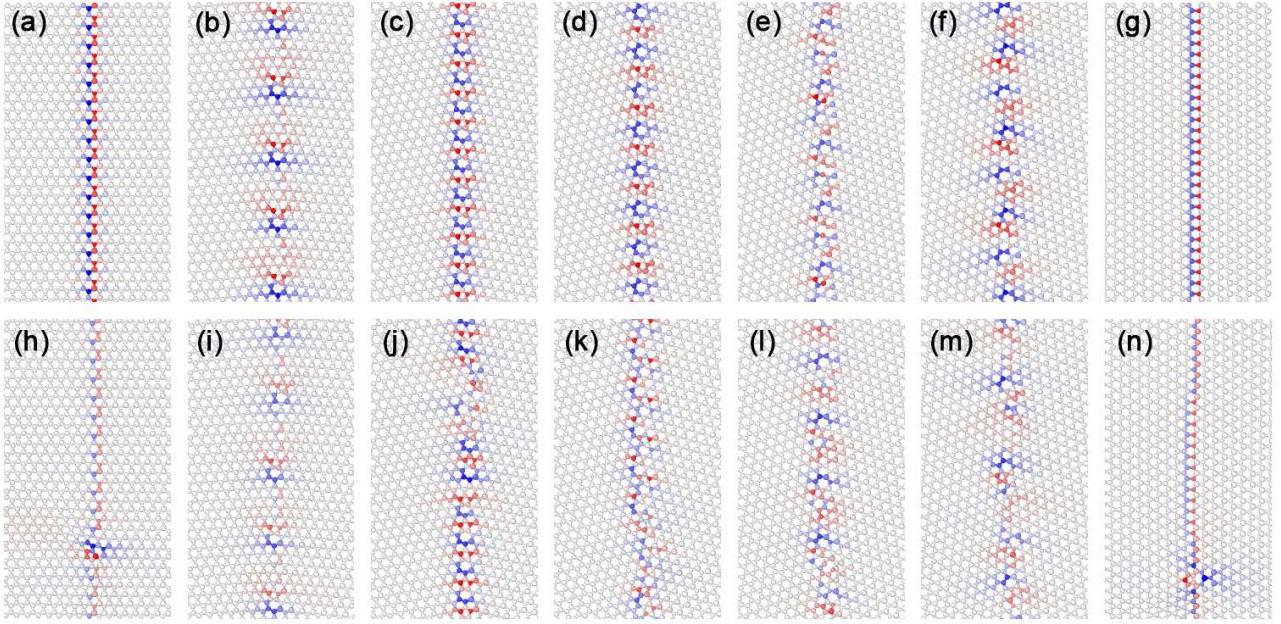}
\caption{The distribution of stress in graphene/h-BN GBs. (a)-(g): The series of lattice-matched-GBs with increasing tilt angles; (h)-(n): The series of lattice-mismatched GBs with increasing tilt angles. For both the upper and lower panels, the tilt angles $2\theta$ from left to right are respectively $0 ^\circ$, $9.43 ^\circ$, $21.79 ^\circ$, $32.20 ^\circ$, $42.10 ^\circ$, $46.83 ^\circ$, and $60 ^\circ$. The sign and magnitude of the stress values ($xx$-component) are represented by the blue and red colors and their density.}
\label{figure:stress_distribution}
\end{center}
\end{figure*}

Third, in the case of $2\theta=60^{\circ}$, the lattice-mismatched GB has a larger thermal conductance than the lattice-matched GB with C-B bonds. This means that, in this particular case, the interfacial stress plays a larger role than the interfacial defects in hindering heat transport across the GB. Figure \ref{figure:stress_distribution} shows the per-atom stress (to be specific, the $xx$-component of the stress tensor is shown) distribution around the GBs. The stress distribution is quite uniform in lattice-matched GBs but is mainly focused on the defects in lattice-mismatched GBs. For a given tilt angle, the average stress in lattice-matched GBs is larger than that in lattice-mismatched GBs. Because both defects and stress field can affect phonon transport, the thermal conductance in the lattice-matched GBs is not necessarily larger than that in the lattice-mismatched GBs. Liu \textit{et al.} \cite{liu2016nl} found that in the case of $2\theta=60^{\circ}$ (zigzag oriented GB), the thermal conductance is always larger in the lattice-mismatched case, no matter if the GB is formed by C-N or C-B bonds. However, we find that the lattice-matched GB with C-N bonds has a larger thermal conductance than the lattice-mismatched GB with $2\theta=60^{\circ}$. We do not know the exact origin of this difference, but we note that the calculations by Liu \textit{et al.} \cite{liu2016nl} were based on a heat current expression as implemented in the LAMMPS package\cite{plimpton1995jcp}, which is not equivalent to that implemented in the GPUMD package \cite{fan2015prb,fan2017cpc} for many-body potentials such as the Tersoff potential used in the present work. Based on our results in Fig. \ref{figure:quantum_conductance}, we can say that there is no clear order in the relative magnitude of the thermal conductance between lattice-matched and lattice-mismatched GBs, and the relative difference is quite small (of the order of $10\%$).

\begin{figure}[htb]
\begin{center}
\includegraphics[width=\columnwidth]{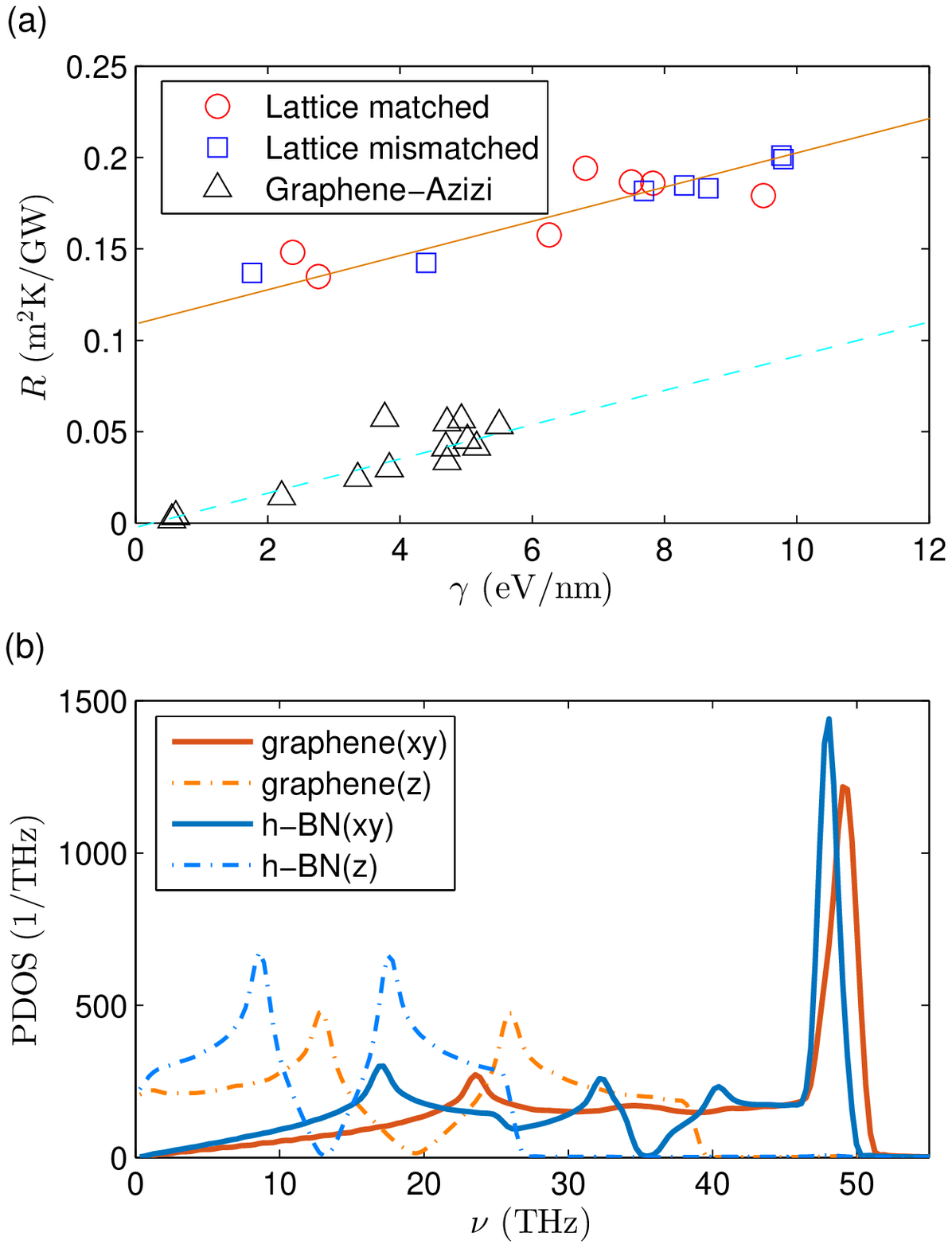}
\caption{(a) Interfacial thermal resistance $R$ as a function of the line tension $\gamma$ for heterogeneous graphene/h-BN GBs and homogeneous graphene GBs \cite{azizi2017cb}. The solid and dashed lines are linear fits to the symbols. (b) Phonon density of states (PDOS) as a function of the phonon frequency $\nu$ for in-plane ($xy$ plane) and out-of-plane ($z$ direction) phonon modes.}
\label{figure:resistance-dos}
\end{center}
\end{figure}

We note that the dependence of the interfacial thermal conductance on the tilt angle in the heterogeneous graphene/h-BN GBs studied here is not as strong as in the case of homogeneous GBs such as in the case of graphene\cite{azizi2017cb}. To better understand this, we show in Fig. \ref{figure:resistance-dos}(a) the interfacial thermal resistance (inverse of thermal conductance) $R$ as a function of the grain boundary energy (line tension) $\gamma$. Previous results for graphene GBs are also shown for comparison. We see that in both graphene/h-BN and graphene GBs, $R$ has roughly a linear dependence on $\gamma$ and the slopes are almost identical. However, when $\gamma\to 0$, $R\to 0$ in graphene but $R$ approaches a finite value in graphene/h-BN GBs. This indicates a clear difference between heterogeneous and homogeneous GBs. Compared to homogeneous GBs, there is an extra mechanism of suppressing the phonon transmission in heterogeneous GBs due to the intrinsic mismatch of the phonon density of states (PDOS) between the two materials, as shown in Fig. \ref{figure:resistance-dos}(b). This results in a finite $R$ in the limit of $\gamma \to 0$.

We have mentioned that flatness can help to enhance the phonon transmission (hence the interfacial thermal conductance) in the graphene/h-BN GBs. This is related to the different PDOS in the in-plane and the out-out-plane modes as shown in Fig. \ref{figure:resistance-dos}(b). When the GB is flat, there will be higher overlap of the PDOS from both sides of the GB, because the in-plane and out-of-plane directions are the same on both sides of the GB. When the GB is non-flat (corrugated), the in-plane and out-of-plane directions are not the same on both sides of the GB, and the overlap of PDOS from both sides will be decreased, resulting in reduced phonon transmission.

\subsection{Weak thermal rectification effects}

We have so far considered heat transport from the graphene sheet to the h-BN sheet only. If we switch the positions of the heat source and sink, heat will flow from the h-BN sheet to the graphene sheet. If the thermal conductance values in the two cases above are different, we can say there is a thermal rectification effect \cite{terraneo2002prl,li2004prl}.

\begin{figure}[htb]
\begin{center}
\includegraphics[width=\columnwidth]{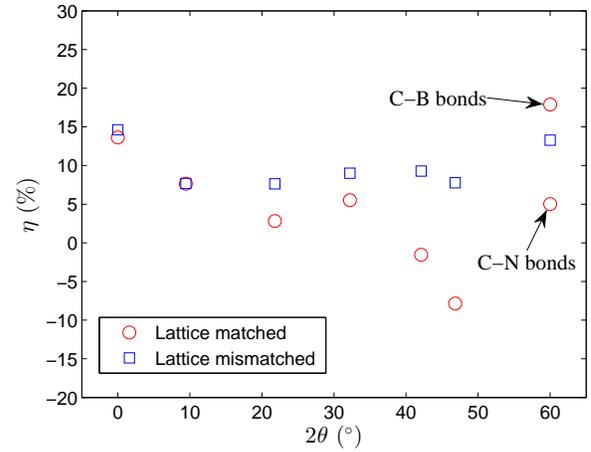}
\caption{Thermal rectification ratio as define in Eq. (equation:tr) for all the GBs studied here.}
\label{figure:rectification}
\end{center}
\end{figure}

To quantitatively study thermal rectification, we define the following rectification ratio,
\begin{equation}
\label{equation:tr}
\eta = \frac{G^{\rm q}_{\rm{BN}\to \rm{G}} - G^{\rm q}_{\rm{G}\to \rm{BN}}}{ \min \{G^{\rm q}_{\rm{G}\to \rm{BN}}, G^{\rm q}_{\rm{BN}\to \rm{G}} \} } \times 100\%.
\end{equation}
Therefore, a positive $\eta$ means that the thermal conductance from h-BN to graphene is higher than the opposite way. Figure \ref{figure:rectification} shows the rectification ratio in all the GBs. For lattice-mismatched GBs, the conductance is always higher when heat flows from the h-BN sheet to the graphene sheet, regardless of the tilt angle; for lattice-matched GBs, this is also the case except for $2\theta=42.10^{\circ}$ and $2\theta=46.83^{\circ}$. In both cases, the rectification ratio is less than $20\%$. These rectification ratios are too small to be applicable for thermal diodes. \cite{li2012rmp}

\section{Summary and Conclusions}

In summary, we have employed a multiscale modeling approach combining PFC model and MD simulations to study heat transport across graphene/h-BN GBs. Using the PFC model, we first constructed a series of realistic bicrystalline samples with different tilt angles. These samples were then used as input to atomistic non-equilibrium MD simulations of heat transport across the GBs. In particular, we considered both lattice-matched and lattice-mismatched conditions and found that lattice match or mismatch does not play a major role in determining the interfacial thermal conductance. Instead, the interfacial thermal conductance is more sensitive to the tilt angle. Furthermore, we found that there exists non-negligible but small thermal rectification in the graphene/h-BN GBs. These results should be useful in the use of lateral graphene/h-BN heterostructures in nanophonoic applications.

\noindent \textbf{Availability of data}

The data that support the findings of this study are partly openly available in a Zenodo repository \cite{zenodo_link} and partly available from the corresponding author upon reasonable request.

\noindent \textbf{Declaration of conflicts}

The authors have no conflicts to disclose.

\begin{acknowledgments}
 This work was supported by the National Key Research and Development Program of China under Grant Nos. 2016YFB0700500 and 2018YFB0704300, the National Natural Science Foundation of China under Grant No. 11974059, the Science Foundation from Education Department of Liaoning Province under Grant No. LQ2020008, and the Academy of Finland through its QTF Centre of Excellence Programme under project No. 312298. We acknowledge the computational resources provided by Aalto Science-IT project and Finland's IT Center for Science (CSC).
\end{acknowledgments}


\begin{thebibliography}{52}%
\makeatletter
\providecommand \@ifxundefined [1]{%
 \@ifx{#1\undefined}
}%
\providecommand \@ifnum [1]{%
 \ifnum #1\expandafter \@firstoftwo
 \else \expandafter \@secondoftwo
 \fi
}%
\providecommand \@ifx [1]{%
 \ifx #1\expandafter \@firstoftwo
 \else \expandafter \@secondoftwo
 \fi
}%
\providecommand \natexlab [1]{#1}%
\providecommand \enquote  [1]{``#1''}%
\providecommand \bibnamefont  [1]{#1}%
\providecommand \bibfnamefont [1]{#1}%
\providecommand \citenamefont [1]{#1}%
\providecommand \href@noop [0]{\@secondoftwo}%
\providecommand \href [0]{\begingroup \@sanitize@url \@href}%
\providecommand \@href[1]{\@@startlink{#1}\@@href}%
\providecommand \@@href[1]{\endgroup#1\@@endlink}%
\providecommand \@sanitize@url [0]{\catcode `\\12\catcode `\$12\catcode
  `\&12\catcode `\#12\catcode `\^12\catcode `\_12\catcode `\%12\relax}%
\providecommand \@@startlink[1]{}%
\providecommand \@@endlink[0]{}%
\providecommand \url  [0]{\begingroup\@sanitize@url \@url }%
\providecommand \@url [1]{\endgroup\@href {#1}{\urlprefix }}%
\providecommand \urlprefix  [0]{URL }%
\providecommand \Eprint [0]{\href }%
\providecommand \doibase [0]{http://dx.doi.org/}%
\providecommand \selectlanguage [0]{\@gobble}%
\providecommand \bibinfo  [0]{\@secondoftwo}%
\providecommand \bibfield  [0]{\@secondoftwo}%
\providecommand \translation [1]{[#1]}%
\providecommand \BibitemOpen [0]{}%
\providecommand \bibitemStop [0]{}%
\providecommand \bibitemNoStop [0]{.\EOS\space}%
\providecommand \EOS [0]{\spacefactor3000\relax}%
\providecommand \BibitemShut  [1]{\csname bibitem#1\endcsname}%
\let\auto@bib@innerbib\@empty
\bibitem [{\citenamefont {Novoselov}\ \emph {et~al.}(2016)\citenamefont
  {Novoselov}, \citenamefont {Mishchenko}, \citenamefont {Carvalho},\ and\
  \citenamefont {Castro~Neto}}]{Novoselov2016science}%
  \BibitemOpen
  \bibfield  {author} {\bibinfo {author} {\bibfnamefont {K.~S.}\ \bibnamefont
  {Novoselov}}, \bibinfo {author} {\bibfnamefont {A.}~\bibnamefont
  {Mishchenko}}, \bibinfo {author} {\bibfnamefont {A.}~\bibnamefont
  {Carvalho}}, \ and\ \bibinfo {author} {\bibfnamefont {A.~H.}\ \bibnamefont
  {Castro~Neto}},\ }\href {\doibase 10.1126/science.aac9439} {\bibfield
  {journal} {\bibinfo  {journal} {Science}\ }\textbf {\bibinfo {volume} {353}}
  (\bibinfo {year} {2016}),\ 10.1126/science.aac9439}\BibitemShut {NoStop}%
\bibitem [{\citenamefont {Ci}\ \emph {et~al.}(2010)\citenamefont {Ci},
  \citenamefont {Song}, \citenamefont {Jin}, \citenamefont {Jariwala},
  \citenamefont {Wu}, \citenamefont {Li}, \citenamefont {Srivastava},
  \citenamefont {Wang}, \citenamefont {Storr}, \citenamefont {Balicas} \emph
  {et~al.}}]{ci2010atomic}%
  \BibitemOpen
  \bibfield  {author} {\bibinfo {author} {\bibfnamefont {L.}~\bibnamefont
  {Ci}}, \bibinfo {author} {\bibfnamefont {L.}~\bibnamefont {Song}}, \bibinfo
  {author} {\bibfnamefont {C.}~\bibnamefont {Jin}}, \bibinfo {author}
  {\bibfnamefont {D.}~\bibnamefont {Jariwala}}, \bibinfo {author}
  {\bibfnamefont {D.}~\bibnamefont {Wu}}, \bibinfo {author} {\bibfnamefont
  {Y.}~\bibnamefont {Li}}, \bibinfo {author} {\bibfnamefont {A.}~\bibnamefont
  {Srivastava}}, \bibinfo {author} {\bibfnamefont {Z.}~\bibnamefont {Wang}},
  \bibinfo {author} {\bibfnamefont {K.}~\bibnamefont {Storr}}, \bibinfo
  {author} {\bibfnamefont {L.}~\bibnamefont {Balicas}},  \emph {et~al.},\
  }\href {\doibase 10.1038/nmat2711} {\bibfield  {journal} {\bibinfo  {journal}
  {Nature materials}\ }\textbf {\bibinfo {volume} {9}},\ \bibinfo {pages} {430}
  (\bibinfo {year} {2010})}\BibitemShut {NoStop}%
\bibitem [{\citenamefont {Levendorf}\ \emph {et~al.}(2012)\citenamefont
  {Levendorf}, \citenamefont {Kim}, \citenamefont {Brown}, \citenamefont
  {Huang}, \citenamefont {Havener}, \citenamefont {Muller},\ and\ \citenamefont
  {Park}}]{Levendorf2012nature}%
  \BibitemOpen
  \bibfield  {author} {\bibinfo {author} {\bibfnamefont {M.~P.}\ \bibnamefont
  {Levendorf}}, \bibinfo {author} {\bibfnamefont {C.-J.}\ \bibnamefont {Kim}},
  \bibinfo {author} {\bibfnamefont {L.}~\bibnamefont {Brown}}, \bibinfo
  {author} {\bibfnamefont {P.~Y.}\ \bibnamefont {Huang}}, \bibinfo {author}
  {\bibfnamefont {R.}~\bibnamefont {Havener}}, \bibinfo {author} {\bibfnamefont
  {D.~A.}\ \bibnamefont {Muller}}, \ and\ \bibinfo {author} {\bibfnamefont
  {J.}~\bibnamefont {Park}},\ }\href {\doibase 10.1038/nature11408} {\bibfield
  {journal} {\bibinfo  {journal} {Nature}\ }\textbf {\bibinfo {volume} {488}},\
  \bibinfo {pages} {627 } (\bibinfo {year} {2012})}\BibitemShut {NoStop}%
\bibitem [{\citenamefont {Liu}\ \emph {et~al.}(2013)\citenamefont {Liu},
  \citenamefont {Ma}, \citenamefont {Shi}, \citenamefont {Zhou}, \citenamefont
  {Gong}, \citenamefont {Lei}, \citenamefont {Yang}, \citenamefont {Zhang},
  \citenamefont {Yu}, \citenamefont {Hackenberg} \emph
  {et~al.}}]{liu2013plane}%
  \BibitemOpen
  \bibfield  {author} {\bibinfo {author} {\bibfnamefont {Z.}~\bibnamefont
  {Liu}}, \bibinfo {author} {\bibfnamefont {L.}~\bibnamefont {Ma}}, \bibinfo
  {author} {\bibfnamefont {G.}~\bibnamefont {Shi}}, \bibinfo {author}
  {\bibfnamefont {W.}~\bibnamefont {Zhou}}, \bibinfo {author} {\bibfnamefont
  {Y.}~\bibnamefont {Gong}}, \bibinfo {author} {\bibfnamefont {S.}~\bibnamefont
  {Lei}}, \bibinfo {author} {\bibfnamefont {X.}~\bibnamefont {Yang}}, \bibinfo
  {author} {\bibfnamefont {J.}~\bibnamefont {Zhang}}, \bibinfo {author}
  {\bibfnamefont {J.}~\bibnamefont {Yu}}, \bibinfo {author} {\bibfnamefont
  {K.~P.}\ \bibnamefont {Hackenberg}},  \emph {et~al.},\ }\href {\doibase
  10.1038/nnano.2012.256} {\bibfield  {journal} {\bibinfo  {journal} {Nature
  nanotechnology}\ }\textbf {\bibinfo {volume} {8}},\ \bibinfo {pages} {119}
  (\bibinfo {year} {2013})}\BibitemShut {NoStop}%
\bibitem [{\citenamefont {Han}\ \emph {et~al.}(2013)\citenamefont {Han},
  \citenamefont {Rodríguez-Manzo}, \citenamefont {Lee}, \citenamefont
  {Kybert}, \citenamefont {Lerner}, \citenamefont {Qi}, \citenamefont
  {Dattoli}, \citenamefont {Rappe}, \citenamefont {Drndic},\ and\ \citenamefont
  {Johnson}}]{Han2013acsnano}%
  \BibitemOpen
  \bibfield  {author} {\bibinfo {author} {\bibfnamefont {G.~H.}\ \bibnamefont
  {Han}}, \bibinfo {author} {\bibfnamefont {J.~A.}\ \bibnamefont
  {Rodríguez-Manzo}}, \bibinfo {author} {\bibfnamefont {C.-W.}\ \bibnamefont
  {Lee}}, \bibinfo {author} {\bibfnamefont {N.~J.}\ \bibnamefont {Kybert}},
  \bibinfo {author} {\bibfnamefont {M.~B.}\ \bibnamefont {Lerner}}, \bibinfo
  {author} {\bibfnamefont {Z.~J.}\ \bibnamefont {Qi}}, \bibinfo {author}
  {\bibfnamefont {E.~N.}\ \bibnamefont {Dattoli}}, \bibinfo {author}
  {\bibfnamefont {A.~M.}\ \bibnamefont {Rappe}}, \bibinfo {author}
  {\bibfnamefont {M.}~\bibnamefont {Drndic}}, \ and\ \bibinfo {author}
  {\bibfnamefont {A.~T.~C.}\ \bibnamefont {Johnson}},\ }\href {\doibase
  10.1021/nn404331f} {\bibfield  {journal} {\bibinfo  {journal} {ACS Nano}\
  }\textbf {\bibinfo {volume} {7}},\ \bibinfo {pages} {10129} (\bibinfo {year}
  {2013})}\BibitemShut {NoStop}%
\bibitem [{\citenamefont {Gao}\ \emph {et~al.}(2013)\citenamefont {Gao},
  \citenamefont {Zhang}, \citenamefont {Chen}, \citenamefont {Li},
  \citenamefont {Liu}, \citenamefont {Gao}, \citenamefont {Ma}, \citenamefont
  {Chen}, \citenamefont {Cheng}, \citenamefont {Qiu}, \citenamefont {Duan},\
  and\ \citenamefont {Liu}}]{gao2013nl}%
  \BibitemOpen
  \bibfield  {author} {\bibinfo {author} {\bibfnamefont {Y.}~\bibnamefont
  {Gao}}, \bibinfo {author} {\bibfnamefont {Y.}~\bibnamefont {Zhang}}, \bibinfo
  {author} {\bibfnamefont {P.}~\bibnamefont {Chen}}, \bibinfo {author}
  {\bibfnamefont {Y.}~\bibnamefont {Li}}, \bibinfo {author} {\bibfnamefont
  {M.}~\bibnamefont {Liu}}, \bibinfo {author} {\bibfnamefont {T.}~\bibnamefont
  {Gao}}, \bibinfo {author} {\bibfnamefont {D.}~\bibnamefont {Ma}}, \bibinfo
  {author} {\bibfnamefont {Y.}~\bibnamefont {Chen}}, \bibinfo {author}
  {\bibfnamefont {Z.}~\bibnamefont {Cheng}}, \bibinfo {author} {\bibfnamefont
  {X.}~\bibnamefont {Qiu}}, \bibinfo {author} {\bibfnamefont {W.}~\bibnamefont
  {Duan}}, \ and\ \bibinfo {author} {\bibfnamefont {Z.}~\bibnamefont {Liu}},\
  }\href {\doibase 10.1021/nl4021123} {\bibfield  {journal} {\bibinfo
  {journal} {Nano Letters}\ }\textbf {\bibinfo {volume} {13}},\ \bibinfo
  {pages} {3439} (\bibinfo {year} {2013})}\BibitemShut {NoStop}%
\bibitem [{\citenamefont {Liu}\ \emph {et~al.}(2014)\citenamefont {Liu},
  \citenamefont {Li}, \citenamefont {Chen}, \citenamefont {Sun}, \citenamefont
  {Ma}, \citenamefont {Li}, \citenamefont {Gao}, \citenamefont {Gao},
  \citenamefont {Cheng}, \citenamefont {Qiu}, \citenamefont {Fang},
  \citenamefont {Zhang},\ and\ \citenamefont {Liu}}]{liu2014nl}%
  \BibitemOpen
  \bibfield  {author} {\bibinfo {author} {\bibfnamefont {M.}~\bibnamefont
  {Liu}}, \bibinfo {author} {\bibfnamefont {Y.}~\bibnamefont {Li}}, \bibinfo
  {author} {\bibfnamefont {P.}~\bibnamefont {Chen}}, \bibinfo {author}
  {\bibfnamefont {J.}~\bibnamefont {Sun}}, \bibinfo {author} {\bibfnamefont
  {D.}~\bibnamefont {Ma}}, \bibinfo {author} {\bibfnamefont {Q.}~\bibnamefont
  {Li}}, \bibinfo {author} {\bibfnamefont {T.}~\bibnamefont {Gao}}, \bibinfo
  {author} {\bibfnamefont {Y.}~\bibnamefont {Gao}}, \bibinfo {author}
  {\bibfnamefont {Z.}~\bibnamefont {Cheng}}, \bibinfo {author} {\bibfnamefont
  {X.}~\bibnamefont {Qiu}}, \bibinfo {author} {\bibfnamefont {Y.}~\bibnamefont
  {Fang}}, \bibinfo {author} {\bibfnamefont {Y.}~\bibnamefont {Zhang}}, \ and\
  \bibinfo {author} {\bibfnamefont {Z.}~\bibnamefont {Liu}},\ }\href {\doibase
  10.1021/nl502780u} {\bibfield  {journal} {\bibinfo  {journal} {Nano Letters}\
  }\textbf {\bibinfo {volume} {14}},\ \bibinfo {pages} {6342} (\bibinfo {year}
  {2014})}\BibitemShut {NoStop}%
\bibitem [{\citenamefont {Ling}\ \emph {et~al.}(2016)\citenamefont {Ling},
  \citenamefont {Lin}, \citenamefont {Ma}, \citenamefont {Wang}, \citenamefont
  {Song}, \citenamefont {Yu}, \citenamefont {Huang}, \citenamefont {Fang},
  \citenamefont {Zhang}, \citenamefont {Hsu}, \citenamefont {Bie},
  \citenamefont {Lee}, \citenamefont {Zhu}, \citenamefont {Wu}, \citenamefont
  {Li}, \citenamefont {Jarillo-Herrero}, \citenamefont {Dresselhaus},
  \citenamefont {Palacios},\ and\ \citenamefont {Kong}}]{Ling2016am}%
  \BibitemOpen
  \bibfield  {author} {\bibinfo {author} {\bibfnamefont {X.}~\bibnamefont
  {Ling}}, \bibinfo {author} {\bibfnamefont {Y.}~\bibnamefont {Lin}}, \bibinfo
  {author} {\bibfnamefont {Q.}~\bibnamefont {Ma}}, \bibinfo {author}
  {\bibfnamefont {Z.}~\bibnamefont {Wang}}, \bibinfo {author} {\bibfnamefont
  {Y.}~\bibnamefont {Song}}, \bibinfo {author} {\bibfnamefont {L.}~\bibnamefont
  {Yu}}, \bibinfo {author} {\bibfnamefont {S.}~\bibnamefont {Huang}}, \bibinfo
  {author} {\bibfnamefont {W.}~\bibnamefont {Fang}}, \bibinfo {author}
  {\bibfnamefont {X.}~\bibnamefont {Zhang}}, \bibinfo {author} {\bibfnamefont
  {A.~L.}\ \bibnamefont {Hsu}}, \bibinfo {author} {\bibfnamefont
  {Y.}~\bibnamefont {Bie}}, \bibinfo {author} {\bibfnamefont {Y.-H.}\
  \bibnamefont {Lee}}, \bibinfo {author} {\bibfnamefont {Y.}~\bibnamefont
  {Zhu}}, \bibinfo {author} {\bibfnamefont {L.}~\bibnamefont {Wu}}, \bibinfo
  {author} {\bibfnamefont {J.}~\bibnamefont {Li}}, \bibinfo {author}
  {\bibfnamefont {P.}~\bibnamefont {Jarillo-Herrero}}, \bibinfo {author}
  {\bibfnamefont {M.}~\bibnamefont {Dresselhaus}}, \bibinfo {author}
  {\bibfnamefont {T.}~\bibnamefont {Palacios}}, \ and\ \bibinfo {author}
  {\bibfnamefont {J.}~\bibnamefont {Kong}},\ }\href {\doibase
  https://doi.org/10.1002/adma.201505070} {\bibfield  {journal} {\bibinfo
  {journal} {Advanced Materials}\ }\textbf {\bibinfo {volume} {28}},\ \bibinfo
  {pages} {2322} (\bibinfo {year} {2016})}\BibitemShut {NoStop}%
\bibitem [{\citenamefont {Volz}\ \emph {et~al.}(2016)\citenamefont {Volz},
  \citenamefont {Ordonez-Miranda}, \citenamefont {Shchepetov}, \citenamefont
  {Prunnila}, \citenamefont {Ahopelto}, \citenamefont {Pezeril}, \citenamefont
  {Vaudel}, \citenamefont {Gusev}, \citenamefont {Ruello}, \citenamefont
  {Weig}, \citenamefont {Schubert}, \citenamefont {Hettich}, \citenamefont
  {Grossman}, \citenamefont {Dekorsy}, \citenamefont {Alzina}, \citenamefont
  {Graczykowski}, \citenamefont {Chavez-Angel}, \citenamefont
  {Sebastian~Reparaz}, \citenamefont {Wagner}, \citenamefont
  {Sotomayor-Torres}, \citenamefont {Xiong}, \citenamefont {Neogi},\ and\
  \citenamefont {Donadio}}]{volz2016epjp}%
  \BibitemOpen
  \bibfield  {author} {\bibinfo {author} {\bibfnamefont {S.}~\bibnamefont
  {Volz}}, \bibinfo {author} {\bibfnamefont {J.}~\bibnamefont
  {Ordonez-Miranda}}, \bibinfo {author} {\bibfnamefont {A.}~\bibnamefont
  {Shchepetov}}, \bibinfo {author} {\bibfnamefont {M.}~\bibnamefont
  {Prunnila}}, \bibinfo {author} {\bibfnamefont {J.}~\bibnamefont {Ahopelto}},
  \bibinfo {author} {\bibfnamefont {T.}~\bibnamefont {Pezeril}}, \bibinfo
  {author} {\bibfnamefont {G.}~\bibnamefont {Vaudel}}, \bibinfo {author}
  {\bibfnamefont {V.}~\bibnamefont {Gusev}}, \bibinfo {author} {\bibfnamefont
  {P.}~\bibnamefont {Ruello}}, \bibinfo {author} {\bibfnamefont {E.~M.}\
  \bibnamefont {Weig}}, \bibinfo {author} {\bibfnamefont {M.}~\bibnamefont
  {Schubert}}, \bibinfo {author} {\bibfnamefont {M.}~\bibnamefont {Hettich}},
  \bibinfo {author} {\bibfnamefont {M.}~\bibnamefont {Grossman}}, \bibinfo
  {author} {\bibfnamefont {T.}~\bibnamefont {Dekorsy}}, \bibinfo {author}
  {\bibfnamefont {F.}~\bibnamefont {Alzina}}, \bibinfo {author} {\bibfnamefont
  {B.}~\bibnamefont {Graczykowski}}, \bibinfo {author} {\bibfnamefont
  {E.}~\bibnamefont {Chavez-Angel}}, \bibinfo {author} {\bibfnamefont
  {J.}~\bibnamefont {Sebastian~Reparaz}}, \bibinfo {author} {\bibfnamefont
  {M.~R.}\ \bibnamefont {Wagner}}, \bibinfo {author} {\bibfnamefont {C.~M.}\
  \bibnamefont {Sotomayor-Torres}}, \bibinfo {author} {\bibfnamefont
  {S.}~\bibnamefont {Xiong}}, \bibinfo {author} {\bibfnamefont
  {S.}~\bibnamefont {Neogi}}, \ and\ \bibinfo {author} {\bibfnamefont
  {D.}~\bibnamefont {Donadio}},\ }\href {\doibase 10.1140/epjb/e2015-60727-7}
  {\bibfield  {journal} {\bibinfo  {journal} {The European Physical Journal B}\
  }\textbf {\bibinfo {volume} {89}},\ \bibinfo {pages} {15} (\bibinfo {year}
  {2016})}\BibitemShut {NoStop}%
\bibitem [{\citenamefont {Ong}, \citenamefont {Zhang},\ and\ \citenamefont
  {Zhang}(2016)}]{ong2016prb}%
  \BibitemOpen
  \bibfield  {author} {\bibinfo {author} {\bibfnamefont {Z.-Y.}\ \bibnamefont
  {Ong}}, \bibinfo {author} {\bibfnamefont {G.}~\bibnamefont {Zhang}}, \ and\
  \bibinfo {author} {\bibfnamefont {Y.-W.}\ \bibnamefont {Zhang}},\ }\href
  {\doibase 10.1103/PhysRevB.93.075406} {\bibfield  {journal} {\bibinfo
  {journal} {Phys. Rev. B}\ }\textbf {\bibinfo {volume} {93}},\ \bibinfo
  {pages} {075406} (\bibinfo {year} {2016})}\BibitemShut {NoStop}%
\bibitem [{\citenamefont {Liu}, \citenamefont {Zhang},\ and\ \citenamefont
  {Zhang}(2016)}]{liu2016nl}%
  \BibitemOpen
  \bibfield  {author} {\bibinfo {author} {\bibfnamefont {X.}~\bibnamefont
  {Liu}}, \bibinfo {author} {\bibfnamefont {G.}~\bibnamefont {Zhang}}, \ and\
  \bibinfo {author} {\bibfnamefont {Y.-W.}\ \bibnamefont {Zhang}},\ }\href
  {\doibase 10.1021/acs.nanolett.6b01565} {\bibfield  {journal} {\bibinfo
  {journal} {Nano Letters}\ }\textbf {\bibinfo {volume} {16}},\ \bibinfo
  {pages} {4954} (\bibinfo {year} {2016})}\BibitemShut {NoStop}%
\bibitem [{\citenamefont {Liang}\ \emph {et~al.}(2020)\citenamefont {Liang},
  \citenamefont {Zhou}, \citenamefont {Zhang}, \citenamefont {Yuan},\ and\
  \citenamefont {Yang}}]{liang2020ijhmt}%
  \BibitemOpen
  \bibfield  {author} {\bibinfo {author} {\bibfnamefont {T.}~\bibnamefont
  {Liang}}, \bibinfo {author} {\bibfnamefont {M.}~\bibnamefont {Zhou}},
  \bibinfo {author} {\bibfnamefont {P.}~\bibnamefont {Zhang}}, \bibinfo
  {author} {\bibfnamefont {P.}~\bibnamefont {Yuan}}, \ and\ \bibinfo {author}
  {\bibfnamefont {D.}~\bibnamefont {Yang}},\ }\href {\doibase
  https://doi.org/10.1016/j.ijheatmasstransfer.2020.119395} {\bibfield
  {journal} {\bibinfo  {journal} {International Journal of Heat and Mass
  Transfer}\ }\textbf {\bibinfo {volume} {151}},\ \bibinfo {pages} {119395}
  (\bibinfo {year} {2020})}\BibitemShut {NoStop}%
\bibitem [{\citenamefont {Li}\ \emph {et~al.}(2018)\citenamefont {Li},
  \citenamefont {Zheng}, \citenamefont {Duan}, \citenamefont {Zhang},
  \citenamefont {Huang},\ and\ \citenamefont {Zhou}}]{li2018jpcc}%
  \BibitemOpen
  \bibfield  {author} {\bibinfo {author} {\bibfnamefont {M.}~\bibnamefont
  {Li}}, \bibinfo {author} {\bibfnamefont {B.}~\bibnamefont {Zheng}}, \bibinfo
  {author} {\bibfnamefont {K.}~\bibnamefont {Duan}}, \bibinfo {author}
  {\bibfnamefont {Y.}~\bibnamefont {Zhang}}, \bibinfo {author} {\bibfnamefont
  {Z.}~\bibnamefont {Huang}}, \ and\ \bibinfo {author} {\bibfnamefont
  {H.}~\bibnamefont {Zhou}},\ }\href {\doibase 10.1021/acs.jpcc.8b02750}
  {\bibfield  {journal} {\bibinfo  {journal} {The Journal of Physical Chemistry
  C}\ }\textbf {\bibinfo {volume} {122}},\ \bibinfo {pages} {14945} (\bibinfo
  {year} {2018})}\BibitemShut {NoStop}%
\bibitem [{\citenamefont {Song}\ \emph {et~al.}(2020)\citenamefont {Song},
  \citenamefont {Xu}, \citenamefont {He}, \citenamefont {Cai}, \citenamefont
  {Bai}, \citenamefont {Miao},\ and\ \citenamefont {Wang}}]{song2020pccp}%
  \BibitemOpen
  \bibfield  {author} {\bibinfo {author} {\bibfnamefont {J.}~\bibnamefont
  {Song}}, \bibinfo {author} {\bibfnamefont {Z.}~\bibnamefont {Xu}}, \bibinfo
  {author} {\bibfnamefont {X.}~\bibnamefont {He}}, \bibinfo {author}
  {\bibfnamefont {C.}~\bibnamefont {Cai}}, \bibinfo {author} {\bibfnamefont
  {Y.}~\bibnamefont {Bai}}, \bibinfo {author} {\bibfnamefont {L.}~\bibnamefont
  {Miao}}, \ and\ \bibinfo {author} {\bibfnamefont {R.}~\bibnamefont {Wang}},\
  }\href {\doibase 10.1039/D0CP01727B} {\bibfield  {journal} {\bibinfo
  {journal} {Phys. Chem. Chem. Phys.}\ }\textbf {\bibinfo {volume} {22}},\
  \bibinfo {pages} {11537} (\bibinfo {year} {2020})}\BibitemShut {NoStop}%
\bibitem [{\citenamefont {Wu}\ and\ \citenamefont {Han}(2021)}]{wu2021jpcc}%
  \BibitemOpen
  \bibfield  {author} {\bibinfo {author} {\bibfnamefont {X.}~\bibnamefont
  {Wu}}\ and\ \bibinfo {author} {\bibfnamefont {Q.}~\bibnamefont {Han}},\
  }\href {\doibase 10.1021/acs.jpcc.0c10387} {\bibfield  {journal} {\bibinfo
  {journal} {The Journal of Physical Chemistry C}\ }\textbf {\bibinfo {volume}
  {125}},\ \bibinfo {pages} {2748} (\bibinfo {year} {2021})}\BibitemShut
  {NoStop}%
\bibitem [{\citenamefont {Ni}\ \emph {et~al.}(2019)\citenamefont {Ni},
  \citenamefont {Zhang}, \citenamefont {Hu}, \citenamefont {Wang},
  \citenamefont {Volz},\ and\ \citenamefont {Xiong}}]{ni2019ijhmt}%
  \BibitemOpen
  \bibfield  {author} {\bibinfo {author} {\bibfnamefont {Y.}~\bibnamefont
  {Ni}}, \bibinfo {author} {\bibfnamefont {H.}~\bibnamefont {Zhang}}, \bibinfo
  {author} {\bibfnamefont {S.}~\bibnamefont {Hu}}, \bibinfo {author}
  {\bibfnamefont {H.}~\bibnamefont {Wang}}, \bibinfo {author} {\bibfnamefont
  {S.}~\bibnamefont {Volz}}, \ and\ \bibinfo {author} {\bibfnamefont
  {S.}~\bibnamefont {Xiong}},\ }\href {\doibase
  https://doi.org/10.1016/j.ijheatmasstransfer.2019.118608} {\bibfield
  {journal} {\bibinfo  {journal} {International Journal of Heat and Mass
  Transfer}\ }\textbf {\bibinfo {volume} {144}},\ \bibinfo {pages} {118608}
  (\bibinfo {year} {2019})}\BibitemShut {NoStop}%
\bibitem [{\citenamefont {Hirvonen}\ \emph {et~al.}(2019)\citenamefont
  {Hirvonen}, \citenamefont {Heinonen}, \citenamefont {Dong}, \citenamefont
  {Fan}, \citenamefont {Elder},\ and\ \citenamefont
  {Ala-Nissila}}]{hirvonen2019prb}%
  \BibitemOpen
  \bibfield  {author} {\bibinfo {author} {\bibfnamefont {P.}~\bibnamefont
  {Hirvonen}}, \bibinfo {author} {\bibfnamefont {V.}~\bibnamefont {Heinonen}},
  \bibinfo {author} {\bibfnamefont {H.}~\bibnamefont {Dong}}, \bibinfo {author}
  {\bibfnamefont {Z.}~\bibnamefont {Fan}}, \bibinfo {author} {\bibfnamefont
  {K.~R.}\ \bibnamefont {Elder}}, \ and\ \bibinfo {author} {\bibfnamefont
  {T.}~\bibnamefont {Ala-Nissila}},\ }\href {\doibase
  10.1103/PhysRevB.100.165412} {\bibfield  {journal} {\bibinfo  {journal}
  {Phys. Rev. B}\ }\textbf {\bibinfo {volume} {100}},\ \bibinfo {pages}
  {165412} (\bibinfo {year} {2019})}\BibitemShut {NoStop}%
\bibitem [{\citenamefont {Elder}\ \emph {et~al.}(2002)\citenamefont {Elder},
  \citenamefont {Katakowski}, \citenamefont {Haataja},\ and\ \citenamefont
  {Grant}}]{elder2002prl}%
  \BibitemOpen
  \bibfield  {author} {\bibinfo {author} {\bibfnamefont {K.~R.}\ \bibnamefont
  {Elder}}, \bibinfo {author} {\bibfnamefont {M.}~\bibnamefont {Katakowski}},
  \bibinfo {author} {\bibfnamefont {M.}~\bibnamefont {Haataja}}, \ and\
  \bibinfo {author} {\bibfnamefont {M.}~\bibnamefont {Grant}},\ }\href
  {\doibase 10.1103/PhysRevLett.88.245701} {\bibfield  {journal} {\bibinfo
  {journal} {Phys. Rev. Lett.}\ }\textbf {\bibinfo {volume} {88}},\ \bibinfo
  {pages} {245701} (\bibinfo {year} {2002})}\BibitemShut {NoStop}%
\bibitem [{\citenamefont {Elder}\ and\ \citenamefont
  {Grant}(2004)}]{elder2004pre}%
  \BibitemOpen
  \bibfield  {author} {\bibinfo {author} {\bibfnamefont {K.~R.}\ \bibnamefont
  {Elder}}\ and\ \bibinfo {author} {\bibfnamefont {M.}~\bibnamefont {Grant}},\
  }\href {\doibase 10.1103/PhysRevE.70.051605} {\bibfield  {journal} {\bibinfo
  {journal} {Phys. Rev. E}\ }\textbf {\bibinfo {volume} {70}},\ \bibinfo
  {pages} {051605} (\bibinfo {year} {2004})}\BibitemShut {NoStop}%
\bibitem [{\citenamefont {Hirvonen}\ \emph {et~al.}(2016)\citenamefont
  {Hirvonen}, \citenamefont {Ervasti}, \citenamefont {Fan}, \citenamefont
  {Jalalvand}, \citenamefont {Seymour}, \citenamefont {Vaez~Allaei},
  \citenamefont {Provatas}, \citenamefont {Harju}, \citenamefont {Elder},\ and\
  \citenamefont {Ala-Nissila}}]{hirvonen2016prb}%
  \BibitemOpen
  \bibfield  {author} {\bibinfo {author} {\bibfnamefont {P.}~\bibnamefont
  {Hirvonen}}, \bibinfo {author} {\bibfnamefont {M.~M.}\ \bibnamefont
  {Ervasti}}, \bibinfo {author} {\bibfnamefont {Z.}~\bibnamefont {Fan}},
  \bibinfo {author} {\bibfnamefont {M.}~\bibnamefont {Jalalvand}}, \bibinfo
  {author} {\bibfnamefont {M.}~\bibnamefont {Seymour}}, \bibinfo {author}
  {\bibfnamefont {S.~M.}\ \bibnamefont {Vaez~Allaei}}, \bibinfo {author}
  {\bibfnamefont {N.}~\bibnamefont {Provatas}}, \bibinfo {author}
  {\bibfnamefont {A.}~\bibnamefont {Harju}}, \bibinfo {author} {\bibfnamefont
  {K.~R.}\ \bibnamefont {Elder}}, \ and\ \bibinfo {author} {\bibfnamefont
  {T.}~\bibnamefont {Ala-Nissila}},\ }\href {\doibase
  10.1103/PhysRevB.94.035414} {\bibfield  {journal} {\bibinfo  {journal} {Phys.
  Rev. B}\ }\textbf {\bibinfo {volume} {94}},\ \bibinfo {pages} {035414}
  (\bibinfo {year} {2016})}\BibitemShut {NoStop}%
\bibitem [{\citenamefont {Taha}\ \emph {et~al.}(2017)\citenamefont {Taha},
  \citenamefont {Mkhonta}, \citenamefont {Elder},\ and\ \citenamefont
  {Huang}}]{taha2017prl}%
  \BibitemOpen
  \bibfield  {author} {\bibinfo {author} {\bibfnamefont {D.}~\bibnamefont
  {Taha}}, \bibinfo {author} {\bibfnamefont {S.~K.}\ \bibnamefont {Mkhonta}},
  \bibinfo {author} {\bibfnamefont {K.~R.}\ \bibnamefont {Elder}}, \ and\
  \bibinfo {author} {\bibfnamefont {Z.-F.}\ \bibnamefont {Huang}},\ }\href
  {\doibase 10.1103/PhysRevLett.118.255501} {\bibfield  {journal} {\bibinfo
  {journal} {Phys. Rev. Lett.}\ }\textbf {\bibinfo {volume} {118}},\ \bibinfo
  {pages} {255501} (\bibinfo {year} {2017})}\BibitemShut {NoStop}%
\bibitem [{\citenamefont {Fan}\ \emph {et~al.}(2017{\natexlab{a}})\citenamefont
  {Fan}, \citenamefont {Hirvonen}, \citenamefont {Pereira}, \citenamefont
  {Ervasti}, \citenamefont {Elder}, \citenamefont {Donadio}, \citenamefont
  {Harju},\ and\ \citenamefont {Ala-Nissila}}]{fan2017nl}%
  \BibitemOpen
  \bibfield  {author} {\bibinfo {author} {\bibfnamefont {Z.}~\bibnamefont
  {Fan}}, \bibinfo {author} {\bibfnamefont {P.}~\bibnamefont {Hirvonen}},
  \bibinfo {author} {\bibfnamefont {L.~F.~C.}\ \bibnamefont {Pereira}},
  \bibinfo {author} {\bibfnamefont {M.~M.}\ \bibnamefont {Ervasti}}, \bibinfo
  {author} {\bibfnamefont {K.~R.}\ \bibnamefont {Elder}}, \bibinfo {author}
  {\bibfnamefont {D.}~\bibnamefont {Donadio}}, \bibinfo {author} {\bibfnamefont
  {A.}~\bibnamefont {Harju}}, \ and\ \bibinfo {author} {\bibfnamefont
  {T.}~\bibnamefont {Ala-Nissila}},\ }\href {\doibase
  10.1021/acs.nanolett.7b01742} {\bibfield  {journal} {\bibinfo  {journal}
  {Nano Letters}\ }\textbf {\bibinfo {volume} {17}},\ \bibinfo {pages} {5919}
  (\bibinfo {year} {2017}{\natexlab{a}})}\BibitemShut {NoStop}%
\bibitem [{\citenamefont {Azizi}\ \emph {et~al.}(2017)\citenamefont {Azizi},
  \citenamefont {Hirvonen}, \citenamefont {Fan}, \citenamefont {Harju},
  \citenamefont {Elder}, \citenamefont {Ala-Nissila},\ and\ \citenamefont
  {Allaei}}]{azizi2017cb}%
  \BibitemOpen
  \bibfield  {author} {\bibinfo {author} {\bibfnamefont {K.}~\bibnamefont
  {Azizi}}, \bibinfo {author} {\bibfnamefont {P.}~\bibnamefont {Hirvonen}},
  \bibinfo {author} {\bibfnamefont {Z.}~\bibnamefont {Fan}}, \bibinfo {author}
  {\bibfnamefont {A.}~\bibnamefont {Harju}}, \bibinfo {author} {\bibfnamefont
  {K.~R.}\ \bibnamefont {Elder}}, \bibinfo {author} {\bibfnamefont
  {T.}~\bibnamefont {Ala-Nissila}}, \ and\ \bibinfo {author} {\bibfnamefont
  {S.~M.~V.}\ \bibnamefont {Allaei}},\ }\href {\doibase
  https://doi.org/10.1016/j.carbon.2017.09.059} {\bibfield  {journal} {\bibinfo
   {journal} {Carbon}\ }\textbf {\bibinfo {volume} {125}},\ \bibinfo {pages}
  {384 } (\bibinfo {year} {2017})}\BibitemShut {NoStop}%
\bibitem [{\citenamefont {Dong}\ \emph {et~al.}(2018)\citenamefont {Dong},
  \citenamefont {Hirvonen}, \citenamefont {Fan},\ and\ \citenamefont
  {Ala-Nissila}}]{dong2018pccp}%
  \BibitemOpen
  \bibfield  {author} {\bibinfo {author} {\bibfnamefont {H.}~\bibnamefont
  {Dong}}, \bibinfo {author} {\bibfnamefont {P.}~\bibnamefont {Hirvonen}},
  \bibinfo {author} {\bibfnamefont {Z.}~\bibnamefont {Fan}}, \ and\ \bibinfo
  {author} {\bibfnamefont {T.}~\bibnamefont {Ala-Nissila}},\ }\href {\doibase
  10.1039/C8CP05159C} {\bibfield  {journal} {\bibinfo  {journal} {Phys. Chem.
  Chem. Phys.}\ }\textbf {\bibinfo {volume} {20}},\ \bibinfo {pages} {24602}
  (\bibinfo {year} {2018})}\BibitemShut {NoStop}%
\bibitem [{\citenamefont {Provatas}\ and\ \citenamefont
  {Elder}(2011)}]{ref-pf-book}%
  \BibitemOpen
  \bibfield  {author} {\bibinfo {author} {\bibfnamefont {N.}~\bibnamefont
  {Provatas}}\ and\ \bibinfo {author} {\bibfnamefont {K.}~\bibnamefont
  {Elder}},\ }\href@noop {} {\emph {\bibinfo {title} {Phase-field methods in
  materials science and engineering}}}\ (\bibinfo  {publisher} {John Wiley \&
  Sons},\ \bibinfo {year} {2011})\BibitemShut {NoStop}%
\bibitem [{\citenamefont {Hirvonen}(2019)}]{ref-hirvonen-thesis}%
  \BibitemOpen
  \bibfield  {author} {\bibinfo {author} {\bibfnamefont {P.}~\bibnamefont
  {Hirvonen}},\ }\emph {\bibinfo {title} {Phase field crystal modeling of
  two-dimensional materials}},\ \href@noop {} {Ph.D. thesis},\ \bibinfo
  {school} {Aalto University} (\bibinfo {year} {2019}),\ \bibinfo {note} {see
  the related section 4.3.2}\BibitemShut {NoStop}%
\bibitem [{\citenamefont {Fan}, \citenamefont {Siro},\ and\ \citenamefont
  {Harju}(2013)}]{fan2013cpc}%
  \BibitemOpen
  \bibfield  {author} {\bibinfo {author} {\bibfnamefont {Z.}~\bibnamefont
  {Fan}}, \bibinfo {author} {\bibfnamefont {T.}~\bibnamefont {Siro}}, \ and\
  \bibinfo {author} {\bibfnamefont {A.}~\bibnamefont {Harju}},\ }\href
  {\doibase https://doi.org/10.1016/j.cpc.2013.01.008} {\bibfield  {journal}
  {\bibinfo  {journal} {Computer Physics Communications}\ }\textbf {\bibinfo
  {volume} {184}},\ \bibinfo {pages} {1414 } (\bibinfo {year}
  {2013})}\BibitemShut {NoStop}%
\bibitem [{\citenamefont {Fan}\ \emph {et~al.}(2017{\natexlab{b}})\citenamefont
  {Fan}, \citenamefont {Chen}, \citenamefont {Vierimaa},\ and\ \citenamefont
  {Harju}}]{fan2017cpc}%
  \BibitemOpen
  \bibfield  {author} {\bibinfo {author} {\bibfnamefont {Z.}~\bibnamefont
  {Fan}}, \bibinfo {author} {\bibfnamefont {W.}~\bibnamefont {Chen}}, \bibinfo
  {author} {\bibfnamefont {V.}~\bibnamefont {Vierimaa}}, \ and\ \bibinfo
  {author} {\bibfnamefont {A.}~\bibnamefont {Harju}},\ }\href {\doibase
  https://doi.org/10.1016/j.cpc.2017.05.003} {\bibfield  {journal} {\bibinfo
  {journal} {Computer Physics Communications}\ }\textbf {\bibinfo {volume}
  {218}},\ \bibinfo {pages} {10 } (\bibinfo {year}
  {2017}{\natexlab{b}})}\BibitemShut {NoStop}%
\bibitem [{\citenamefont {Tersoff}(1989)}]{Tersoff1989prb}%
  \BibitemOpen
  \bibfield  {author} {\bibinfo {author} {\bibfnamefont {J.}~\bibnamefont
  {Tersoff}},\ }\href {\doibase 10.1103/PhysRevB.39.5566} {\bibfield  {journal}
  {\bibinfo  {journal} {Phys. Rev. B}\ }\textbf {\bibinfo {volume} {39}},\
  \bibinfo {pages} {5566} (\bibinfo {year} {1989})}\BibitemShut {NoStop}%
\bibitem [{\citenamefont {Kinaci}\ \emph {et~al.}(2012)\citenamefont {Kinaci},
  \citenamefont {Haskins}, \citenamefont {Sevik},\ and\ \citenamefont
  {Cagin}}]{kinaci2012prb}%
  \BibitemOpen
  \bibfield  {author} {\bibinfo {author} {\bibfnamefont {A.}~\bibnamefont
  {Kinaci}}, \bibinfo {author} {\bibfnamefont {J.~B.}\ \bibnamefont {Haskins}},
  \bibinfo {author} {\bibfnamefont {C.}~\bibnamefont {Sevik}}, \ and\ \bibinfo
  {author} {\bibfnamefont {T.}~\bibnamefont {Cagin}},\ }\href {\doibase
  10.1103/PhysRevB.86.115410} {\bibfield  {journal} {\bibinfo  {journal} {Phys.
  Rev. B}\ }\textbf {\bibinfo {volume} {86}},\ \bibinfo {pages} {115410}
  (\bibinfo {year} {2012})}\BibitemShut {NoStop}%
\bibitem [{\citenamefont {Berendsen}\ \emph {et~al.}(1984)\citenamefont
  {Berendsen}, \citenamefont {Postma}, \citenamefont {van Gunsteren},
  \citenamefont {DiNola},\ and\ \citenamefont {Haak}}]{Berendsen1984jcp}%
  \BibitemOpen
  \bibfield  {author} {\bibinfo {author} {\bibfnamefont {H.~J.~C.}\
  \bibnamefont {Berendsen}}, \bibinfo {author} {\bibfnamefont {J.~P.~M.}\
  \bibnamefont {Postma}}, \bibinfo {author} {\bibfnamefont {W.~F.}\
  \bibnamefont {van Gunsteren}}, \bibinfo {author} {\bibfnamefont
  {A.}~\bibnamefont {DiNola}}, \ and\ \bibinfo {author} {\bibfnamefont {J.~R.}\
  \bibnamefont {Haak}},\ }\href {\doibase 10.1063/1.448118} {\bibfield
  {journal} {\bibinfo  {journal} {The Journal of Chemical Physics}\ }\textbf
  {\bibinfo {volume} {81}},\ \bibinfo {pages} {3684} (\bibinfo {year}
  {1984})}\BibitemShut {NoStop}%
\bibitem [{\citenamefont {Li}\ \emph {et~al.}(2019)\citenamefont {Li},
  \citenamefont {Xiong}, \citenamefont {Sievers}, \citenamefont {Hu},
  \citenamefont {Fan}, \citenamefont {Wei}, \citenamefont {Bao}, \citenamefont
  {Chen}, \citenamefont {Donadio},\ and\ \citenamefont
  {Ala-Nissila}}]{li2019jcp}%
  \BibitemOpen
  \bibfield  {author} {\bibinfo {author} {\bibfnamefont {Z.}~\bibnamefont
  {Li}}, \bibinfo {author} {\bibfnamefont {S.}~\bibnamefont {Xiong}}, \bibinfo
  {author} {\bibfnamefont {C.}~\bibnamefont {Sievers}}, \bibinfo {author}
  {\bibfnamefont {Y.}~\bibnamefont {Hu}}, \bibinfo {author} {\bibfnamefont
  {Z.}~\bibnamefont {Fan}}, \bibinfo {author} {\bibfnamefont {N.}~\bibnamefont
  {Wei}}, \bibinfo {author} {\bibfnamefont {H.}~\bibnamefont {Bao}}, \bibinfo
  {author} {\bibfnamefont {S.}~\bibnamefont {Chen}}, \bibinfo {author}
  {\bibfnamefont {D.}~\bibnamefont {Donadio}}, \ and\ \bibinfo {author}
  {\bibfnamefont {T.}~\bibnamefont {Ala-Nissila}},\ }\href {\doibase
  10.1063/1.5132543} {\bibfield  {journal} {\bibinfo  {journal} {The Journal of
  Chemical Physics}\ }\textbf {\bibinfo {volume} {151}},\ \bibinfo {pages}
  {234105} (\bibinfo {year} {2019})}\BibitemShut {NoStop}%
\bibitem [{\citenamefont {Hu}\ \emph {et~al.}(2020)\citenamefont {Hu},
  \citenamefont {Feng}, \citenamefont {Gu}, \citenamefont {Fan}, \citenamefont
  {Wang}, \citenamefont {Lundstrom}, \citenamefont {Shrestha},\ and\
  \citenamefont {Bao}}]{hu2020prb}%
  \BibitemOpen
  \bibfield  {author} {\bibinfo {author} {\bibfnamefont {Y.}~\bibnamefont
  {Hu}}, \bibinfo {author} {\bibfnamefont {T.}~\bibnamefont {Feng}}, \bibinfo
  {author} {\bibfnamefont {X.}~\bibnamefont {Gu}}, \bibinfo {author}
  {\bibfnamefont {Z.}~\bibnamefont {Fan}}, \bibinfo {author} {\bibfnamefont
  {X.}~\bibnamefont {Wang}}, \bibinfo {author} {\bibfnamefont {M.}~\bibnamefont
  {Lundstrom}}, \bibinfo {author} {\bibfnamefont {S.~S.}\ \bibnamefont
  {Shrestha}}, \ and\ \bibinfo {author} {\bibfnamefont {H.}~\bibnamefont
  {Bao}},\ }\href {\doibase 10.1103/PhysRevB.101.155308} {\bibfield  {journal}
  {\bibinfo  {journal} {Phys. Rev. B}\ }\textbf {\bibinfo {volume} {101}},\
  \bibinfo {pages} {155308} (\bibinfo {year} {2020})}\BibitemShut {NoStop}%
\bibitem [{\citenamefont {Bussi}\ and\ \citenamefont
  {Parrinello}(2007)}]{Bussi2007pre}%
  \BibitemOpen
  \bibfield  {author} {\bibinfo {author} {\bibfnamefont {G.}~\bibnamefont
  {Bussi}}\ and\ \bibinfo {author} {\bibfnamefont {M.}~\bibnamefont
  {Parrinello}},\ }\href {\doibase 10.1103/PhysRevE.75.056707} {\bibfield
  {journal} {\bibinfo  {journal} {Phys. Rev. E}\ }\textbf {\bibinfo {volume}
  {75}},\ \bibinfo {pages} {056707} (\bibinfo {year} {2007})}\BibitemShut
  {NoStop}%
\bibitem [{\citenamefont {Fan}\ \emph {et~al.}(2015)\citenamefont {Fan},
  \citenamefont {Pereira}, \citenamefont {Wang}, \citenamefont {Zheng},
  \citenamefont {Donadio},\ and\ \citenamefont {Harju}}]{fan2015prb}%
  \BibitemOpen
  \bibfield  {author} {\bibinfo {author} {\bibfnamefont {Z.}~\bibnamefont
  {Fan}}, \bibinfo {author} {\bibfnamefont {L.~F.~C.}\ \bibnamefont {Pereira}},
  \bibinfo {author} {\bibfnamefont {H.-Q.}\ \bibnamefont {Wang}}, \bibinfo
  {author} {\bibfnamefont {J.-C.}\ \bibnamefont {Zheng}}, \bibinfo {author}
  {\bibfnamefont {D.}~\bibnamefont {Donadio}}, \ and\ \bibinfo {author}
  {\bibfnamefont {A.}~\bibnamefont {Harju}},\ }\href {\doibase
  10.1103/PhysRevB.92.094301} {\bibfield  {journal} {\bibinfo  {journal} {Phys.
  Rev. B}\ }\textbf {\bibinfo {volume} {92}},\ \bibinfo {pages} {094301}
  (\bibinfo {year} {2015})}\BibitemShut {NoStop}%
\bibitem [{\citenamefont {Fan}\ \emph {et~al.}(2019)\citenamefont {Fan},
  \citenamefont {Dong}, \citenamefont {Harju},\ and\ \citenamefont
  {Ala-Nissila}}]{fan2019prb}%
  \BibitemOpen
  \bibfield  {author} {\bibinfo {author} {\bibfnamefont {Z.}~\bibnamefont
  {Fan}}, \bibinfo {author} {\bibfnamefont {H.}~\bibnamefont {Dong}}, \bibinfo
  {author} {\bibfnamefont {A.}~\bibnamefont {Harju}}, \ and\ \bibinfo {author}
  {\bibfnamefont {T.}~\bibnamefont {Ala-Nissila}},\ }\href {\doibase
  10.1103/PhysRevB.99.064308} {\bibfield  {journal} {\bibinfo  {journal} {Phys.
  Rev. B}\ }\textbf {\bibinfo {volume} {99}},\ \bibinfo {pages} {064308}
  (\bibinfo {year} {2019})}\BibitemShut {NoStop}%
\bibitem [{\citenamefont {Gabourie}\ \emph {et~al.}(2021)\citenamefont
  {Gabourie}, \citenamefont {Fan}, \citenamefont {Ala-Nissila},\ and\
  \citenamefont {Pop}}]{Gabourie2021prb}%
  \BibitemOpen
  \bibfield  {author} {\bibinfo {author} {\bibfnamefont {A.~J.}\ \bibnamefont
  {Gabourie}}, \bibinfo {author} {\bibfnamefont {Z.}~\bibnamefont {Fan}},
  \bibinfo {author} {\bibfnamefont {T.}~\bibnamefont {Ala-Nissila}}, \ and\
  \bibinfo {author} {\bibfnamefont {E.}~\bibnamefont {Pop}},\ }\href {\doibase
  10.1103/PhysRevB.103.205421} {\bibfield  {journal} {\bibinfo  {journal}
  {Phys. Rev. B}\ }\textbf {\bibinfo {volume} {103}},\ \bibinfo {pages}
  {205421} (\bibinfo {year} {2021})}\BibitemShut {NoStop}%
\bibitem [{\citenamefont {Fan}\ \emph {et~al.}(2017{\natexlab{c}})\citenamefont
  {Fan}, \citenamefont {Pereira}, \citenamefont {Hirvonen}, \citenamefont
  {Ervasti}, \citenamefont {Elder}, \citenamefont {Donadio}, \citenamefont
  {Ala-Nissila},\ and\ \citenamefont {Harju}}]{fan2017prb}%
  \BibitemOpen
  \bibfield  {author} {\bibinfo {author} {\bibfnamefont {Z.}~\bibnamefont
  {Fan}}, \bibinfo {author} {\bibfnamefont {L.~F.~C.}\ \bibnamefont {Pereira}},
  \bibinfo {author} {\bibfnamefont {P.}~\bibnamefont {Hirvonen}}, \bibinfo
  {author} {\bibfnamefont {M.~M.}\ \bibnamefont {Ervasti}}, \bibinfo {author}
  {\bibfnamefont {K.~R.}\ \bibnamefont {Elder}}, \bibinfo {author}
  {\bibfnamefont {D.}~\bibnamefont {Donadio}}, \bibinfo {author} {\bibfnamefont
  {T.}~\bibnamefont {Ala-Nissila}}, \ and\ \bibinfo {author} {\bibfnamefont
  {A.}~\bibnamefont {Harju}},\ }\href {\doibase 10.1103/PhysRevB.95.144309}
  {\bibfield  {journal} {\bibinfo  {journal} {Phys. Rev. B}\ }\textbf {\bibinfo
  {volume} {95}},\ \bibinfo {pages} {144309} (\bibinfo {year}
  {2017}{\natexlab{c}})}\BibitemShut {NoStop}%
\bibitem [{\citenamefont {Xu}\ \emph {et~al.}(2018)\citenamefont {Xu},
  \citenamefont {Fan}, \citenamefont {Zhang}, \citenamefont {Wei},\ and\
  \citenamefont {Ala-Nissila}}]{xu2018msmse}%
  \BibitemOpen
  \bibfield  {author} {\bibinfo {author} {\bibfnamefont {K.}~\bibnamefont
  {Xu}}, \bibinfo {author} {\bibfnamefont {Z.}~\bibnamefont {Fan}}, \bibinfo
  {author} {\bibfnamefont {J.}~\bibnamefont {Zhang}}, \bibinfo {author}
  {\bibfnamefont {N.}~\bibnamefont {Wei}}, \ and\ \bibinfo {author}
  {\bibfnamefont {T.}~\bibnamefont {Ala-Nissila}},\ }\href
  {http://stacks.iop.org/0965-0393/26/i=8/a=085001} {\bibfield  {journal}
  {\bibinfo  {journal} {Modelling and Simulation in Materials Science and
  Engineering}\ }\textbf {\bibinfo {volume} {26}},\ \bibinfo {pages} {085001}
  (\bibinfo {year} {2018})}\BibitemShut {NoStop}%
\bibitem [{\citenamefont {Chalopin}\ and\ \citenamefont
  {Volz}(2013)}]{Chalopin2013apl}%
  \BibitemOpen
  \bibfield  {author} {\bibinfo {author} {\bibfnamefont {Y.}~\bibnamefont
  {Chalopin}}\ and\ \bibinfo {author} {\bibfnamefont {S.}~\bibnamefont
  {Volz}},\ }\href {\doibase 10.1063/1.4816738} {\bibfield  {journal} {\bibinfo
   {journal} {Applied Physics Letters}\ }\textbf {\bibinfo {volume} {103}},\
  \bibinfo {pages} {051602} (\bibinfo {year} {2013})}\BibitemShut {NoStop}%
\bibitem [{\citenamefont {S\"a\"askilahti}\ \emph {et~al.}(2014)\citenamefont
  {S\"a\"askilahti}, \citenamefont {Oksanen}, \citenamefont {Tulkki},\ and\
  \citenamefont {Volz}}]{saaskilahti2014prb}%
  \BibitemOpen
  \bibfield  {author} {\bibinfo {author} {\bibfnamefont {K.}~\bibnamefont
  {S\"a\"askilahti}}, \bibinfo {author} {\bibfnamefont {J.}~\bibnamefont
  {Oksanen}}, \bibinfo {author} {\bibfnamefont {J.}~\bibnamefont {Tulkki}}, \
  and\ \bibinfo {author} {\bibfnamefont {S.}~\bibnamefont {Volz}},\ }\href
  {\doibase 10.1103/PhysRevB.90.134312} {\bibfield  {journal} {\bibinfo
  {journal} {Phys. Rev. B}\ }\textbf {\bibinfo {volume} {90}},\ \bibinfo
  {pages} {134312} (\bibinfo {year} {2014})}\BibitemShut {NoStop}%
\bibitem [{\citenamefont {S\"a\"askilahti}\ \emph {et~al.}(2015)\citenamefont
  {S\"a\"askilahti}, \citenamefont {Oksanen}, \citenamefont {Volz},\ and\
  \citenamefont {Tulkki}}]{saaskilahti2015prb}%
  \BibitemOpen
  \bibfield  {author} {\bibinfo {author} {\bibfnamefont {K.}~\bibnamefont
  {S\"a\"askilahti}}, \bibinfo {author} {\bibfnamefont {J.}~\bibnamefont
  {Oksanen}}, \bibinfo {author} {\bibfnamefont {S.}~\bibnamefont {Volz}}, \
  and\ \bibinfo {author} {\bibfnamefont {J.}~\bibnamefont {Tulkki}},\ }\href
  {\doibase 10.1103/PhysRevB.91.115426} {\bibfield  {journal} {\bibinfo
  {journal} {Phys. Rev. B}\ }\textbf {\bibinfo {volume} {91}},\ \bibinfo
  {pages} {115426} (\bibinfo {year} {2015})}\BibitemShut {NoStop}%
\bibitem [{\citenamefont {Zhou}\ and\ \citenamefont {Hu}(2015)}]{zhou2015prb}%
  \BibitemOpen
  \bibfield  {author} {\bibinfo {author} {\bibfnamefont {Y.}~\bibnamefont
  {Zhou}}\ and\ \bibinfo {author} {\bibfnamefont {M.}~\bibnamefont {Hu}},\
  }\href {\doibase 10.1103/PhysRevB.92.195205} {\bibfield  {journal} {\bibinfo
  {journal} {Phys. Rev. B}\ }\textbf {\bibinfo {volume} {92}},\ \bibinfo
  {pages} {195205} (\bibinfo {year} {2015})}\BibitemShut {NoStop}%
\bibitem [{\citenamefont {Fan}\ \emph {et~al.}(2021)\citenamefont {Fan},
  \citenamefont {Zeng}, \citenamefont {Zhang}, \citenamefont {Wang},
  \citenamefont {Dong}, \citenamefont {Chen},\ and\ \citenamefont
  {Ala-Nissila}}]{fan2021arxiv}%
  \BibitemOpen
  \bibfield  {author} {\bibinfo {author} {\bibfnamefont {Z.}~\bibnamefont
  {Fan}}, \bibinfo {author} {\bibfnamefont {Z.}~\bibnamefont {Zeng}}, \bibinfo
  {author} {\bibfnamefont {C.}~\bibnamefont {Zhang}}, \bibinfo {author}
  {\bibfnamefont {Y.}~\bibnamefont {Wang}}, \bibinfo {author} {\bibfnamefont
  {H.}~\bibnamefont {Dong}}, \bibinfo {author} {\bibfnamefont {Y.}~\bibnamefont
  {Chen}}, \ and\ \bibinfo {author} {\bibfnamefont {T.}~\bibnamefont
  {Ala-Nissila}},\ }\href@noop {} {\enquote {\bibinfo {title} {Neuroevolution
  machine learning potentials: Combining high accuracy and low cost in
  atomistic simulations and application to heat transport},}\ } (\bibinfo
  {year} {2021}),\ \Eprint {http://arxiv.org/abs/2107.08119} {arXiv:2107.08119
  [physics.comp-ph]} \BibitemShut {NoStop}%
\bibitem [{\citenamefont {Lv}\ and\ \citenamefont {Henry}(2016)}]{lv2016njp}%
  \BibitemOpen
  \bibfield  {author} {\bibinfo {author} {\bibfnamefont {W.}~\bibnamefont
  {Lv}}\ and\ \bibinfo {author} {\bibfnamefont {A.}~\bibnamefont {Henry}},\
  }\href {http://stacks.iop.org/1367-2630/18/i=1/a=013028} {\bibfield
  {journal} {\bibinfo  {journal} {New Journal of Physics}\ }\textbf {\bibinfo
  {volume} {18}},\ \bibinfo {pages} {013028} (\bibinfo {year}
  {2016})}\BibitemShut {NoStop}%
\bibitem [{\citenamefont {S\"a\"askilahti}\ \emph {et~al.}(2016)\citenamefont
  {S\"a\"askilahti}, \citenamefont {Oksanen}, \citenamefont {Tulkki},
  \citenamefont {McGaughey},\ and\ \citenamefont {Volz}}]{saaskilahti2016aipa}%
  \BibitemOpen
  \bibfield  {author} {\bibinfo {author} {\bibfnamefont {K.}~\bibnamefont
  {S\"a\"askilahti}}, \bibinfo {author} {\bibfnamefont {J.}~\bibnamefont
  {Oksanen}}, \bibinfo {author} {\bibfnamefont {J.}~\bibnamefont {Tulkki}},
  \bibinfo {author} {\bibfnamefont {A.~J.~H.}\ \bibnamefont {McGaughey}}, \
  and\ \bibinfo {author} {\bibfnamefont {S.}~\bibnamefont {Volz}},\ }\href
  {\doibase 10.1063/1.4968617} {\bibfield  {journal} {\bibinfo  {journal} {AIP
  Advances}\ }\textbf {\bibinfo {volume} {6}},\ \bibinfo {pages} {121904}
  (\bibinfo {year} {2016})}\BibitemShut {NoStop}%
\bibitem [{\citenamefont {Schelling}, \citenamefont {Phillpot},\ and\
  \citenamefont {Keblinski}(2004)}]{Schelling2004jap}%
  \BibitemOpen
  \bibfield  {author} {\bibinfo {author} {\bibfnamefont {P.~K.}\ \bibnamefont
  {Schelling}}, \bibinfo {author} {\bibfnamefont {S.~R.}\ \bibnamefont
  {Phillpot}}, \ and\ \bibinfo {author} {\bibfnamefont {P.}~\bibnamefont
  {Keblinski}},\ }\href {\doibase 10.1063/1.1702100} {\bibfield  {journal}
  {\bibinfo  {journal} {Journal of Applied Physics}\ }\textbf {\bibinfo
  {volume} {95}},\ \bibinfo {pages} {6082} (\bibinfo {year}
  {2004})}\BibitemShut {NoStop}%
\bibitem [{\citenamefont {Plimpton}(1995)}]{plimpton1995jcp}%
  \BibitemOpen
  \bibfield  {author} {\bibinfo {author} {\bibfnamefont {S.}~\bibnamefont
  {Plimpton}},\ }\href {\doibase https://doi.org/10.1006/jcph.1995.1039}
  {\bibfield  {journal} {\bibinfo  {journal} {Journal of Computational
  Physics}\ }\textbf {\bibinfo {volume} {117}},\ \bibinfo {pages} {1 }
  (\bibinfo {year} {1995})}\BibitemShut {NoStop}%
\bibitem [{\citenamefont {Terraneo}, \citenamefont {Peyrard},\ and\
  \citenamefont {Casati}(2002)}]{terraneo2002prl}%
  \BibitemOpen
  \bibfield  {author} {\bibinfo {author} {\bibfnamefont {M.}~\bibnamefont
  {Terraneo}}, \bibinfo {author} {\bibfnamefont {M.}~\bibnamefont {Peyrard}}, \
  and\ \bibinfo {author} {\bibfnamefont {G.}~\bibnamefont {Casati}},\ }\href
  {\doibase 10.1103/PhysRevLett.88.094302} {\bibfield  {journal} {\bibinfo
  {journal} {Phys. Rev. Lett.}\ }\textbf {\bibinfo {volume} {88}},\ \bibinfo
  {pages} {094302} (\bibinfo {year} {2002})}\BibitemShut {NoStop}%
\bibitem [{\citenamefont {Li}, \citenamefont {Wang},\ and\ \citenamefont
  {Casati}(2004)}]{li2004prl}%
  \BibitemOpen
  \bibfield  {author} {\bibinfo {author} {\bibfnamefont {B.}~\bibnamefont
  {Li}}, \bibinfo {author} {\bibfnamefont {L.}~\bibnamefont {Wang}}, \ and\
  \bibinfo {author} {\bibfnamefont {G.}~\bibnamefont {Casati}},\ }\href
  {\doibase 10.1103/PhysRevLett.93.184301} {\bibfield  {journal} {\bibinfo
  {journal} {Phys. Rev. Lett.}\ }\textbf {\bibinfo {volume} {93}},\ \bibinfo
  {pages} {184301} (\bibinfo {year} {2004})}\BibitemShut {NoStop}%
\bibitem [{\citenamefont {Li}\ \emph {et~al.}(2012)\citenamefont {Li},
  \citenamefont {Ren}, \citenamefont {Wang}, \citenamefont {Zhang},
  \citenamefont {H\"anggi},\ and\ \citenamefont {Li}}]{li2012rmp}%
  \BibitemOpen
  \bibfield  {author} {\bibinfo {author} {\bibfnamefont {N.}~\bibnamefont
  {Li}}, \bibinfo {author} {\bibfnamefont {J.}~\bibnamefont {Ren}}, \bibinfo
  {author} {\bibfnamefont {L.}~\bibnamefont {Wang}}, \bibinfo {author}
  {\bibfnamefont {G.}~\bibnamefont {Zhang}}, \bibinfo {author} {\bibfnamefont
  {P.}~\bibnamefont {H\"anggi}}, \ and\ \bibinfo {author} {\bibfnamefont
  {B.}~\bibnamefont {Li}},\ }\href {\doibase 10.1103/RevModPhys.84.1045}
  {\bibfield  {journal} {\bibinfo  {journal} {Rev. Mod. Phys.}\ }\textbf
  {\bibinfo {volume} {84}},\ \bibinfo {pages} {1045} (\bibinfo {year}
  {2012})}\BibitemShut {NoStop}%
\bibitem [{\citenamefont {Fan}(2021)}]{zenodo_link}%
  \BibitemOpen
  \bibfield  {author} {\bibinfo {author} {\bibfnamefont {Z.}~\bibnamefont
  {Fan}},\ }\href {\doibase 10.5281/zenodo.5324321} {\enquote {\bibinfo {title}
  {{graphene/hexagonal-BN grain boundary samples}},}\ } (\bibinfo {year}
  {2021})\BibitemShut {NoStop}%
\end{thebibliography}
\end{document}